\newif\ifShowKeys
\ShowKeystrue
\ShowKeysfalse

\newif\ifshowtikz
\showtikztrue
\showtikzfalse   

\documentclass[11pt]{article}
	\pdfoutput=1
\usepackage[T1]{fontenc}

\usepackage{listings}
\lstnewenvironment{arkady}  
{\lstset{language=C,frame=trbl,basicstyle = \footnotesize \ttfamily , breaklines = true,showstringspaces=false}}{}

\usepackage{froufrou}
\usepackage{datetime}
\usepackage{comment}					

\usepackage{color}
\definecolor{Maroon}{rgb}{0.15,0.33,0.52}
\definecolor{Mahogany}{rgb}{0.65,0.0975,0.0845}
\usepackage[setpagesize=false,pagebackref=false, 
linktocpage, bookmarksopen=true, colorlinks=true, 
linkcolor=Maroon,citecolor=Maroon,urlcolor=Maroon]{hyperref}

\usepackage[parsep]{collref}				

\ifShowKeys \usepackage[notcite]{showkeys} \fi

\usepackage{amsmath, amssymb,amsthm}
\usepackage{stackrel}
\numberwithin{equation}{section}
\usepackage{bm,environ,mathrsfs,array,arydshln}
\usepackage{booktabs,float,slashed}
\usepackage{appendix}
\usepackage[mathcal]{euscript}
\usepackage{tensor} 						
\usepackage{mathabx}
\usepackage[vcentermath]{youngtab}
\usepackage{simpler-wick}

\usepackage{graphicx,epsfig,epic}
\usepackage{subcaption,wrapfig}
\usepackage{tikz}
\usepackage{tikz-feynman} 

\allowdisplaybreaks

\usepackage{framed}						
\definecolor{shadecolor}{rgb}{0.9996078, 0.984314, 0.960784}
\definecolor{framecolor}{rgb}{0,0,0}
\definecolor{TFTitleColor}{RGB}{1,1,1}
\definecolor{TFFrameColor}{RGB}{249	218	181}		
\definecolor{TFFrameColor}{RGB}{230 230 230 }

\newenvironment{frshaded}{%
    \MakeFramed {\FrameRestore}}%
    {\endMakeFramed}

\definecolor{myred}{RGB}{233, 33, 45}

\makeatletter
\DeclareFontFamily{OMX}{MnSymbolE}{}
\DeclareSymbolFont{MnLargeSymbols}{OMX}{MnSymbolE}{m}{n}
\SetSymbolFont{MnLargeSymbols}{bold}{OMX}{MnSymbolE}{b}{n}
\DeclareFontShape{OMX}{MnSymbolE}{m}{n}{
    <-6>  MnSymbolE5
   <6-7>  MnSymbolE6
   <7-8>  MnSymbolE7
   <8-9>  MnSymbolE8
   <9-10> MnSymbolE9
  <10-12> MnSymbolE10
  <12->   MnSymbolE12
}{}
\DeclareFontShape{OMX}{MnSymbolE}{b}{n}{
    <-6>  MnSymbolE-Bold5
   <6-7>  MnSymbolE-Bold6
   <7-8>  MnSymbolE-Bold7
   <8-9>  MnSymbolE-Bold8
   <9-10> MnSymbolE-Bold9
  <10-12> MnSymbolE-Bold10
  <12->   MnSymbolE-Bold12
}{}

\let\llangle\@undefined
\let\rrangle\@undefined
\DeclareMathDelimiter{\llangle}{\mathopen}%
                     {MnLargeSymbols}{'164}{MnLargeSymbols}{'164}
\DeclareMathDelimiter{\rrangle}{\mathclose}%
                     {MnLargeSymbols}{'171}{MnLargeSymbols}{'171}
\makeatother

\usepackage{mdframed}

\definecolor{lightpeach}{RGB}{255, 247, 235}

\newmdenv[
  backgroundcolor=gray!5,
  linecolor=gray!40,
  roundcorner=5pt,
  innerleftmargin=8pt,
  innerrightmargin=8pt,
  innertopmargin=6pt,
  innerbottommargin=6pt,
]{remarkbox}

\newcommand{\bs}{\begin{frshaded}}			
\newcommand{\es}{\end{frshaded}\noindent}

\def\ba#1\ea{\begin{align}#1\end{align}}		        
\newcommand{\be}{\begin{equation}}
\newcommand{\ee}{\end{equation}}
\newcommand{\bea}{\begin{equation} \begin{aligned}} 
\newcommand{\eea}{\end{aligned} \end{equation}}
\newcommand{\mc}{\mathcal }
\newcommand{\wh}{\widehat}
\newcommand{\wt}{\widetilde}

\newcommand{\mk}{\mathfrak}
\newcommand{\la}{\label}
\newcommand{\eps}{\varepsilon}

\newcommand{\lp}{\notag \\ & }

\DeclareMathOperator{\sech}{\text{sech}}

\DeclareMathOperator{\Tr}{\text{Tr}}

\DeclareMathOperator{\LI}{Li}

\newcommand{\cf}{\textit{cf.} }
\newcommand{\ie}{\textit{i.e.} }
\newcommand{\eg}{\textit{e.g.} }

\newcommand{\sql}{\sqrt\l}
\renewcommand{\l}{\lambda}

\newcommand{\ket}[1]{|#1\rangle}

\newcommand{\braket}[2]{\langle #1|#2\rangle}
\newcommand{\mmm}[3]{\langle #1|#2|#3\rangle}

\newcommand{\sG}{{\mathscr G}}

\newcommand{\sC}{{\mathscr C}}

\begin{document}

\begin{titlepage}


\vspace*{15mm}
\begin{center}
{\Large   Three-loop onset of the wormhole length in DSSYK} 
\vspace*{10mm}

E. Alfinito$^{a,b}$ and M. Beccaria$^{a,b}$

\vspace*{4mm}
{\small
	
${}^a$ Universit\`a del Salento, Dipartimento di Matematica e Fisica \textit{Ennio De Giorgi},
\vskip 0.2cm
${}^{b}$ INFN - sezione di Lecce, Via Arnesano, I-73100 Lecce, Italy
\vskip 0.3cm
\vskip 0.2cm {\small E-mail: \texttt{matteo.beccaria@le.infn.it}}
}
\vspace*{0.8cm}
\end{center}

\begin{abstract}
The length of the Einstein-Rosen bridge in sine-dilaton gravity at disk level equals the Krylov spread 
complexity of the dual double-scaled SYK (DSSYK) model, with the double-scaling parameter 
$\l$ controlling the semiclassical expansion.
We compute 
the onset value, $L_{0}$, in the thermofield double state at 
$t=0$ and arbitrary inverse temperature $\beta$, through three loops and in closed form, 
extending the known one-loop result. The quantity $L_{0}$ represents the preparation 
complexity of the initial thermal Hartle-Hawking state. 
In DSSYK, $L_{0}$ is $\l$ times the average chord number 
of the thermal state, providing a microscopic, unambiguous determination of an additive 
constant otherwise fixed only by a choice of holographic renormalization scheme. 
The two-loop contribution follows from a saddle point evaluation of the DSSYK two-point 
function at coincident insertion points, where individually divergent contributions cancel 
only in their sum. At three loops, we bypass the saddle point analysis using an exact 
recursion relation that generates long high-temperature expansions at low cost, fixing and 
testing a proposed Ansatz against many independent data points. 
Finally, in the low-temperature regime each loop order 
contributes one further power of $\beta$, reorganizing the semiclassical series into an 
expansion in the Schwarzian coupling $\l\beta$, whose leading coefficient we check against an 
independent one-loop Schwarzian computation.
The same methods, applied to the length variance and third-order cumulant at $t=0$, give their 
semiclassical expansion through two loops.
\end{abstract}

\end{titlepage}


{\small
\makeatletter
\newcommand*{\toccontents}{\@starttoc{toc}}
\makeatother
\toccontents
}


\vspace{1cm}

\setcounter{footnote}{0}

\section{Introduction}

The Majorana SYK model \cite{Sachdev:1992fk,Sachdev:2010um,Maldacena:2016hyu} describes 
$N$ Majorana fermions with random local interactions coupling sets of $p$ fermions.
The model exhibits quantum chaos with the maximal chaos exponent \cite{Maldacena:2015waa,Cotler:2016fpe}. 
In the large $N$ limit (at fixed $p$), the SYK model admits a low-energy description in terms of the so-called 
Schwarzian quantum mechanics, which is the boundary action of JT gravity on nearly AdS$_{2}$ space 
\cite{Maldacena:2016hyu,Maldacena:2016upp,Jensen:2016pah,Sarosi:2017ykf,Berkooz:2018jqr,Mertens:2017mtv,Engelsoy:2016xyb,Mertens:2022irh}.

From the holographic perspective, a well-studied regime of the SYK model is the so-called double-scaling limit (DSSYK)
where $N$ and $p$ are both taken large with fixed ratio $\l = 2p^{2}/N$ 
\cite{Cotler:2016fpe,Berkooz:2018jqr,Berkooz:2018qkz,Berkooz:2024lgq}. \footnote{
The JT gravity description is reproduced by considering DSSYK at small $\l$ and zooming in on its low energy dynamics.}
At generic coupling $\l$ and finite temperature, the  holographic dual of DSSYK is a two-dimensional sine-dilaton gravity model
\cite{Blommaert:2023opb,Blommaert:2024ymv,Blommaert:2024whf}. \footnote{
Correlation functions can be matched between DSSYK
and gravity results, by introducing in the bulk a non-minimally coupled probe dual to bilocal operators in the 
boundary Schwarzian theory \cite{Blommaert:2024ymv}.}
The seminal papers \cite{Rabinovici:2023yex,Xu:2024gfm,Ambrosini:2024sre} showed that Krylov complexity \cite{Nandy:2024evd,Baiguera:2025dkc,Rabinovici:2025otw}
of the infinite-temperature thermofield double state \cite{Maldacena:2001kr}
on the boundary of AdS$_{2}$ admits a bulk description in JT gravity as the length of the two-sided wormhole.
This result was extended to sine-dilaton gravity context in \cite{Heller:2024ldz}. 
Sine-dilaton theory is dual to DSSYK at disk level \cite{Blommaert:2024ymv}; its extension 
beyond disk level is the subject of ongoing work.\footnote{
In particular the large $N$ expansion of the one-cut matrix model 
\cite{Blommaert:2024whf,Blommaert:2025avl}, and its relation to the genus expansion of the two-matrix model dual to complex Liouville string 
\cite{Collier:2024kmo,Collier:2024lys,Collier:2024kwt,Collier:2024mlg,Collier:2025pbm} is not yet fully 
understood.
} 
The gravity action considered in
\cite{Blommaert:2024ymv,Blommaert:2023opb} reads (we omit boundary terms)
\be
S = \tfrac{1}{2}\int_{\mc M} d^{2}\sigma\sqrt{g}\,(\Phi R+\tfrac{2}{\l}\sin(\l\Phi)),
\ee
where $\Phi$ is the dilaton field. 
The coupling $\l$ plays the role of  a semiclassical expansion parameter: the 
classical limit at $\l\to 0$ corresponds to an AdS$_{2}$ black hole with Hawking (inverse) temperature  $\beta_{\rm BH} = 2\pi/\sin\theta$. 
After holographic renormalization, the classical expression of the wormhole length is
\be
\la{1.2}
L = 2\log\cosh(t\sin(\theta)/2)-2\log\sin\theta+O(\l),
\ee
where the time variable is the two-sided boundary Lorentzian time $t$. The relation between the parameter $\theta$ 
and the microscopic DSSYK inverse temperature $\beta$ is \cite{Blommaert:2024whf} (see also \cite{Lin:2023trc})
\be
\la{1.3}
J\,\beta = \frac{\pi-2\theta}{\sin\theta},
\ee
where the standard choice in the literature is $J=\frac{1}{2}$ and the parameter $J$ may conveniently be absorbed into $\beta$.
In DSSYK it is useful to introduce the parameter 
\be
q \equiv e^{-\l},
\ee
a deformation parameter for the 
oscillator algebra that underlies the double-scaling construction.

A key result of \cite{Heller:2024ldz} is the relation between the expectation value of the quantum 
length $\hat L$ and Krylov complexity. It reads
\be
\la{1.5}
\langle\hat L(\beta, t) \rangle = \l\, C_{\beta}(t),
\ee
where  $C_{\beta}(t)$ is the DSSYK Krylov complexity at Lorentzian time $t$ of the 
Hartle-Hawking state at inverse temperature $\beta$.
On the gravity side, 
the length expectation value was extracted in \cite{Heller:2024ldz} from 
the two-point function of a massive probe given by the insertion of the bilocal operator $e^{-\Delta\hat L}$ 
of conformal weight $\Delta$ \cite{Blommaert:2024ymv}
\be
\la{1.6}
\langle\hat L\rangle = \left[-\partial_{\Delta}\langle e^{-\Delta\hat L}\rangle\right]_{\Delta=0}.
\ee
The two-point function is computed by splitting the disk 
boundary into two parts of length $\tau$ and $\beta-\tau$ with 
\be
\la{1.7}
\tau=\tfrac{1}{2}\beta+it,
\ee
and corresponding to the Euclidean boundary times to which the semiclassical geodesic is anchored. 
In the DSSYK model, the operator $e^{-\Delta \hat L}$ in (\ref{1.6}) corresponds to a matter operator built with a  string of fermions, see \cite{Berkooz:2018jqr} for details.
The classical contribution 
to $\langle e^{-\Delta \hat L}\rangle$ is well-known \cite{Maldacena:2016hyu} and reads
\be
\la{1.8}
\langle e^{-\Delta \hat L}\rangle_{\rm cl} = \bigg(\frac{\sin^{2}\theta}{\sin^{2}(\sin\theta\, \frac{\tau}{2}+\theta)}\bigg)^{\Delta},
\ee
where $\theta, \tau$ are related to $\beta, t$ according to (\ref{1.3}) and (\ref{1.7}).
At one-loop and beyond, one needs to work order by order in small $\Delta$ with the result
\be
\la{1.9}
\langle e^{-\Delta\hat L}\rangle = \langle e^{-\Delta\hat L}\rangle_{\rm cl}\,\bigg[1+\Delta\mc A+\Delta^{2}\mc B+O(\Delta^{3})\bigg],
\ee
where $\mc A, \mc B, \dots$ are power series in $\l$ with coefficients that depend on $\theta$ and $\tau$
\be
\la{1.10}
\mc A = \l\, \mc A_{1}+\l^{2}\mc A_{2}+\cdots, \qquad 
\mc B = \l\, \mc B_{1}+\l^{2}\mc B_{2}+\cdots, \qquad \textit{etc.}
\ee
The first two one-loop coefficients $\mc A_{1}$, $\mc B_{1}$ are \cite{Goel:2023svz,Okuyama:2023bch} 
\bea
\la{1.11}
\mc A_{1}(u,\zeta) &= \frac{1}{4(1+u \tan u)}\bigg[-\frac{(1+u \tan u)^{2}}{\cos^{2}\zeta}+\frac{(1+\zeta\tan \zeta)^{2}}{\cos^{2}u}+\zeta^{2}(\tan^{2}u-\tan^{2}\zeta)
-\frac{1+\zeta\tan\zeta}{1+u \tan u}+1\bigg], \\
\mc B_{1}(u,\zeta) &= -\frac{(\tan\zeta+\tan\zeta(\zeta+u)\tan u+\tan u)(\tan\zeta+\tan u(\tan\zeta(u-\zeta)-1))}{2(1+u\tan u)}.
\eea
The new variables $(u,\zeta)$ are defined in terms of  $(\theta, \tau)$ by 
\be
\la{1.12}
u = \frac{\pi}{2}-\theta, \qquad \zeta = \frac{\pi}{2}-\theta-\frac{\tau}{2}\sin\theta,
\ee
and their relation to the initial variables $(\beta, t)$ is 
\be
\la{1.13}
\beta=\frac{4u}{\cos u}, \qquad \zeta = -\frac{2iu}{\beta}\, t.
\ee
The quantum length expectation value, or Krylov complexity from  (\ref{1.5}), is thus the following extension of (\ref{1.2})
\be
\langle\hat L(\beta, t)\rangle = \l\, C_{\beta}(t) = 2\log\cosh(t\,\sin\theta/2)-2\log\sin\theta-\l\,\mc A_{1}(u,\zeta)+O(\l^{2}).
\ee
In this paper, we are interested in its value at $t=0$
\be
L_{0}(\beta) \equiv \langle \hat L(\beta, 0)\rangle = \l\, C_{\beta}(0).
\ee
This corresponds to $\tau=\beta/2$ and, from (\ref{1.13}), one has to set $\zeta=0$. 
At this special point, the expression $\mc A_{1}(u,0)$ gives
\be
\la{1.16}
L_{0}(\beta) =-2\log\cos u+\frac{u\tan u+(-1+3u^{2})\tan^{2}u+u(-1+u^{2})\tan^{3}u}{4 (1+u \tan u)^2}\,\l+O(\l^{2}).
\ee
In particular, its expansion around the infinite temperature limit $\theta=\pi/2$ is an expansion in small $u$ and we get 
\be
\la{1.17}
L_{0} = u^{2}+\frac{u^{4}}{6}+\frac{2u^{6}}{45}+\frac{17u^{8}}{1260}+\cdots+\bigg(\frac{5u^{4}}{12}-\frac{71u^{6}}{180}+\frac{103u^{8}}{252}+\cdots\bigg)\,\l+O(\l^{2}).
\ee
The quantity $L_{0}$ defined in (\ref{1.16}) is the \emph{onset} of the wormhole length: the 
expectation value of the length operator in the finite-temperature thermofield double state 
\emph{before} any Lorentzian evolution. 
Equivalently, by (\ref{1.5}) and (\ref{2.8}), it is $\l$ times the Krylov spread complexity of the 
state $e^{-\frac{\beta}{2}H}\ket{0}$, \ie the average chord number in the thermal state. Unlike 
$C_{\beta}(t)$, it is not a dynamical observable but a state function of the temperature alone, 
and it plays the role of the $\mathrm{AdS}_{2}$ analogue of the \emph{complexity of formation} 
\cite{Chapman:2016hwi} -- the cost of preparing the entangled two-sided state, as opposed to the 
rate at which its complexity subsequently grows. The same interpretation of this onset, as the 
complexity of Euclidean state preparation distinct from that of subsequent Lorentzian growth, 
was recently emphasized for Krylov complexity itself in \cite{Heller:2024ldz}. It interpolates between two sharply different 
regimes: as $\beta\to 0$ the state degenerates to the zero-chord state and $L_{0}\to 0$, the 
wormhole having zero length; as $\beta\to\infty$ one finds $L_{0}\simeq 2\log(\beta/(2\pi))$, the 
logarithmic lengthening of the near-extremal $\mathrm{AdS}_{2}$ throat.
\footnote{A similar low-temperature logarithmic scaling of 
holographic complexity was found, in the different setting of cold hyperbolic black holes, 
by \cite{Barbon:2015soa}.}

There is a specific reason why $L_{0}$ merits separate treatment. In the bulk, the classical 
solution for the two-sided geodesic length takes the form 
$L(t) = 2\log\cosh(\ldots)+L_{\rm reg}$, where the additive constant $L_{\rm reg}$ is fixed only 
by a choice of holographic renormalization scheme, and is customarily discarded.
In the 
sine-dilaton/DSSYK dictionary this constant is identified with the initial total chord number 
$\ell_{*}(\theta)$ \cite{Aguilar-Gutierrez:2025pqp}, whose matter-free value 
$e^{-\ell_{*}}=\sin^{2}\theta$ reproduces precisely the classical term $-2\log\sin\theta$ in 
(\ref{1.2}) and  (\ref{1.16}). The microscopic side of the correspondence, however, is not scheme dependent at 
all: the length operator $\hat\ell = \l\wh n$ has a discrete spectrum bounded below, and 
the zero-chord state provides a canonical origin. The DSSYK computation therefore {fixes} 
what is an arbitrary constant in the bulk, and the loop corrections computed in this paper are 
unambiguous predictions for it.

This also explains why $L_{0}$ is not available in the existing literature. In the JT gravity 
and matrix-model treatments of the Einstein-Rosen bridge, only the time-{difference} 
$\langle\hat\ell(t)\rangle-\langle\hat\ell(0)\rangle$ is computed, the $t=0$ value being divergent in the 
regulator $\Delta$ and absorbed into a subtraction constant \cite{Iliesiu:2021ari}. On the other 
hand, the methods that give the semiclassical expansion of $C_{\beta}(t)$ to high loop order 
\cite{Alfinito:2026cky} are tailored to the infinite temperature limit $\beta=0$. The present work 
is thus complementary to both: it covers the whole temperature range at the single time $t=0$, 
where the other approaches either subtract the answer or do not apply. Moreover, since 
$\ell_{*}(\theta)$ enters the finite-temperature saddle point solution as the initial data for the 
subsequent evolution, its higher-loop determination is a prerequisite for extending the 
semiclassical expansion of $C_{\beta}(t)$ itself beyond $\beta=0$. \footnote{
The same gap has been noted from the gravity side. The finite-temperature quantum wormhole 
length was set up in \cite{Fu:2025kkh} in precisely the form used here -- as the $\Delta\to0$ 
derivative of the two-sided correlator with boundary states prepared in Euclidean signature, 
equivalent to the Krylov complexity of the zero-chord state -- but only at the level of the 
exact formal expression, and the semiclassical expansion of the resulting length, order by 
order in $\l$, was explicitly left open there, as was the study of the thermofield 
double state in the chord basis.
}

Finally, the point $t=0$ is not merely a convenient specialization; it is the technically most demanding regime of the 
saddle point analysis. In the notation of (\ref{1.12}) it corresponds to $\zeta=0$, \ie a two-point function at coinciding points
where the individual ingredients of the fluctuation determinant develop poles of increasing 
order at each loop -- $1/\eps$ at one loop, up to $1/\eps^{3}$ at two loops -- in the regulator 
$\eps=\l\Delta$, which must be treated as independent of $\l$, as explained in 
Section~\ref{sec:saddle}.
That  these singularities cancel exactly, order by order, is a nontrivial structural feature of the 
symmetric-point correlator, and exhibiting the mechanism explicitly is one of the aims of the 
present paper.

Beyond the specific results just outlined, three points of a more general nature are worth 
emphasizing. First, the resolution of the additive-constant ambiguity described above is an 
instance of a broader phenomenon: a microscopic completion can render unambiguous a quantity 
that is only defined up to a choice of scheme in its effective, gravitational description. 
Second, the semiclassical expansion of $L_{0}$ admits an independent, quantitative test of the 
DSSYK/JT duality itself: in 
Section~\ref{sec:schwarzian-matching} we show that the leading one-loop coefficient, computed 
purely within DSSYK, is exactly reproduced -- both its rational and its transcendental ($\pi$-dependent) parts -- 
by an independent one-loop computation performed entirely within Schwarzian quantum 
mechanics. The same agreement is expected at higher loop order. 
Third, the mechanism responsible for the finiteness of $L_{0}$ at the symmetric 
point is structurally universal: the same obstruction -- individually divergent, 
coincident-point singularities that cancel only in a specific combination -- is encountered 
independently in the DSSYK saddle point calculation of Section~\ref{sec:saddle} and, in an 
entirely different guise, in the exact Schwarzian bilocal correlator of 
Section~\ref{sec:schwarzian-matching}.

\paragraph{Summary of results}

Our main result is the closed form of the semiclassical expansion of $L_{0}$ at arbitrary 
temperature, through three loops. In general, 
\be
\la{1.18}
L_{0} = -2\log\cos u-\sum_{k\ge 1}\mc A_{k}(u,0)\,\l^{k}, \qquad \beta = \frac{4u}{\cos u},
\ee
where the variable $u\in(0,\pi/2)$ introduced in (\ref{1.12}) covers the whole temperature 
range, with $u\to 0$ the infinite temperature limit and $u\to\pi/2$ the low temperature one.
The one-loop coefficient $\mc A_{1}(u,0)$ reproduces the known result of 
\cite{Goel:2023svz,Okuyama:2023bch}, while the two- and three-loop contributions $\mc A_{2}(u,0)$ and $\mc A_{3}(u,0)$, given in 
(\ref{5.53}) and (\ref{6.7}), are new. All three share the structural form
\be
\la{1.19}
\mc A_{k}(u,0) = \frac{1}{(1+u\tan u)^{N_{1}}}\sum_{p=0}^{N_{2}}\mc P_{p}(u^{2})\,
u^{\sigma_{p}}\tan^{p}u, \qquad \sigma_{p} = \begin{cases} 0, & p\ \text{even}, \\ 
1, & p\ \text{odd},\end{cases}
\ee
with $\mc P_{p}$ polynomials and
\be
\la{1.20}
N_{1}=3k-1, \qquad N_{2}=4k-1, \qquad \deg \mc P_{p}\le 2k-1,
\ee
a pattern read off from the explicit one- and two-loop expressions and confirmed at three 
loops. Their expansions around the infinite temperature point begin as
\bea
\la{1.21}
\mc A_{1}(u,0) &= -\frac{5u^{4}}{12}+\frac{71u^{6}}{180}-\frac{103u^{8}}{252}+\cdots, \qquad
\mc A_{2}(u,0) = \frac{5u^{4}}{36}-\frac{131u^{6}}{216}+\frac{41549u^{8}}{30240}+\cdots, \\
\mc A_{3}(u,0) &= -\frac{5u^{4}}{144}+\frac{101u^{6}}{240}-\frac{5639u^{8}}{2880}+\cdots.
\eea
These expansions converge for $|u|<R$ with $R\approx 1.1997$ solution of $R\tanh R=1$, as follows from the closed form (\ref{1.19}).
Note that this is smaller than the maximum physical value $u=\pi/2$ corresponding to zero temperature. Of course, (\ref{1.19})
is smooth for any physical real $0\le u<\pi/2$.

The tool that makes the three-loop determination feasible is an exact three-term recursion 
relation in the chord number, derived in Section~\ref{sec:recursion}. Let $Z_{n}(x)$ be the 
functions resolving the two-point function in the chord number, defined in (\ref{6.3}); 
they are the Euclidean amplitudes connecting the zero-chord state to the $n$-chord state, 
$Z_{n}(x) = (-1)^{n}\sqrt{(q;q)_{n}}\,\mmm{n}{e^{-\beta H}}{0}$, with $x$ the rescaled inverse 
temperature (\ref{3.6}) and $Z_{0}=Z$ the partition function. As a consequence of the 
recurrence obeyed by the continuous $q$-Hermite polynomials, one has
\be
\la{1.22}
Z_{n+1}(x) = 2\,Z_{n}'(x)-(1-q^{n})\,Z_{n-1}(x), \qquad Z_{-1}(x)=0, \quad Z_{0}(x)=Z(x).
\ee
Each step requires only one differentiation and one linear combination, and this generates the 
high temperature expansion of each loop order to very high order at reasonable cost.

Finally, the closed expressions (\ref{1.19}) allow to study the low temperature regime of large $\beta$. One finds $\mc A_{k}(u,0)\simeq K_{k}\tan^{k}u$ as 
$u\to\pi/2$, and since $\tan u\simeq \beta/(2\pi)$ in the same limit, each loop order 
contributes exactly one further power of $\beta$. The semiclassical series therefore 
reorganizes as an expansion in the product $\l\beta$,
\be
\la{1.23}
L_{0} = 2\log\frac{\beta}{2\pi}+\sum_{k\ge1}\wh K_{k}\,(\l\beta)^{k}+\cdots, 
\ee
where the dots stand for terms suppressed by powers of $1/\beta$ at each order, and 
$\wh K_{k} = -K_{k}/(2\pi)^{k}$, explicitly
\be
\la{1.24}
\wh K_{1} = \frac{1}{16}-\frac{1}{4\pi^{2}}, \quad 
\wh K_{2} = -\frac{1}{512}-\frac{1}{32\pi^{4}}+\frac{1}{48\pi^{2}}, \quad
\wh K_{3} = \frac{1}{6144}-\frac{1}{64\pi^{6}}+\frac{1}{192\pi^{4}}-\frac{1}{512\pi^{2}}.
\ee
All the above results are obtained by three independent methods -- a direct $q$-oscillator 
algebra evaluation, a saddle point analysis of the DSSYK two-point function, and the recursion 
(\ref{1.22}) combined with the Ansatz (\ref{1.19}) -- which agree wherever they overlap and 
provide strong mutual cross-checks.

The appearance of $\l\beta$ in (\ref{1.23}) is of course not accidental: it is the Schwarzian coupling of 
the triple-scaling limit relating DSSYK to JT gravity, so that the low temperature expansion of 
$L_{0}$ is a loop expansion in the Schwarzian theory. We return to this in 
Section~\ref{sec:analysis}, where the dictionary is made explicit and 
(\ref{1.18}) is explicitly matched against the Schwarzian two-sided geodesic length at one-loop order. 
Moreover, since $\mc A_{1,2,3}(u,0)$ are exact functions of $u$, our results also determine an 
infinite tower of corrections to (\ref{1.23}), subleading in $1/\beta$ at each loop order, that 
have no counterpart in the strict Schwarzian limit and are captured only by the finite-$\l$, 
exact sine-dilaton dual.

Finally, the $q$-oscillator algebra and the recursion-fixed Ansatz of 
Sections~\ref{sec:small-beta} and~\ref{sec:recursion} extend directly to the variance and 
third-order cumulant of the length, $L_{0}^{(2)}\equiv\langle\hat\ell^{2}\rangle_{c}$ and 
$L_{0}^{(3)}\equiv\langle\hat\ell^{3}\rangle_{c}$, 
both at $t=0$. We compute their semiclassical expansion at two-loop order and their low/high-temperature 
expansions in closed form.

\paragraph{Plan of the paper}
In Section~\ref{sec:L0-def} we recall the chord space formulation of the DSSYK model and 
define $L_{0}$ as the $t=0$ value of the Krylov spread complexity of the zero-chord state 
evolved in Euclidean time. Section~\ref{sec:spectral} introduces the spectral representation 
of the relevant matrix elements and shows that $L_{0}$ is obtained from the $\Delta\to 0$ 
limit of the derivative of the two-point function of a matter operator of dimension $\Delta$. 
Section~\ref{sec:small-beta} derives the high temperature expansion of $L_{0}$ in two 
independent ways, by direct manipulation of the $q$-oscillator algebra and from the small $x$ 
expansion of the two-point function. Section~\ref{sec:saddle} is devoted to the saddle point 
analysis at coincident insertion points: after discussing the classical saddle, we rederive in 
Section~\ref{sec:one-loop} the known one-loop coefficient, paying attention to the mechanism by 
which the singularities in the regulator $\eps$ cancel, and obtain in Section~\ref{two-loop} 
the new two-loop coefficient. Section~\ref{sec:3loop} presents an alternative strategy that 
bypasses the three-loop saddle point calculation: the exact three-term recursion relation 
(\ref{1.22}) is derived in Section~\ref{sec:recursion} and used to generate a long high 
temperature expansion, which fixes the Ansatz (\ref{1.19}) and yields the three-loop 
coefficient. Section~\ref{sec:analysis} analyses the one-, two- and three-loop corrections in the low temperature regime
and their relation to the Schwarzian limit. 
Section~\ref{sec:fluctuations} extends the analysis to the variance and third-order cumulant of the length at $t=0$, through 
two loops. Section~\ref{sec:summary} contains a summary and further comments.

\section{$L_{0}$ from Krylov complexity in the DSSYK model}
\la{sec:L0-def}

The field content of the SYK model is composed of $N$ Majorana fermions $\{\psi_{i}\}_{i=1, \dots, N}$, with canonical anticommutation relations $\{\psi_{i}, \psi_{j}\}=2\delta_{ij}$
and Hamiltonian
\be
H_{\rm SYK} = i^{p/2}\sum_{1\le i_{1}<i_{2}<\cdots<i_{p}\le N} J_{i_{1}\cdots i_{p}}\ \psi_{i_{1}}\cdots \psi_{i_{p}},
\ee
where $J_{i_{1}\cdots i_{p}}$ are Gaussian random couplings with strength
\be
\llangle J_{i_{1}\cdots i_{p}}J_{j_{1}\cdots j_{p}}\rrangle =\frac{N}{2p^{2}}\binom{N}{p}^{-1}J^{2}\delta_{i_{1},j_{1}}\cdots \delta_{i_{p},j_{p}},
\ee
where we denoted by  $\llangle\cdots\rrangle$ the ensemble average and the normalization is as in \cite{Maldacena:2016hyu,Lin:2022rbf,Lin:2023trc}.
The double-scaling (DS)  limit is defined as the combined large $N,p$ limit with fixed $p^{2}/N$ ratio \cite{Erdos:2014zgc,Cotler:2016fpe}
\be
N,p\to\infty, \qquad \l\equiv\frac{2p^{2}}{N},\qquad q \equiv e^{-\l}, 
\ee
where $q$ is the so-called deformation parameter.
In the double-scaling limit, one may introduce an effective Hilbert space of DSSYK \cite{Berkooz:2018jqr,Lin:2022rbf} with an orthonormal basis 
consisting of chord states $\{\ket{n}\}_{n=0,1,2,\dots}$, which are eigenstates of the chord-number operator $\wh n$: $\wh n\ket{n} = n\ket{n}$. 
In this space, one considers the chord-Hamiltonian
\be
\la{2.4}
H = \frac{J}{\sql}(a+a^{\dagger}), 
\ee
where $a$ and $a^{\dagger}$ are $q$-deformed oscillators obeying \cite{Lin:2023trc} the algebra
\be
\la{2.5}
[a, a^{\dagger}]_{q} = a\, a^{\dagger}-q\,a^{\dagger}\,a=1, \qquad [\wh n, a^{\dagger}]=a^{\dagger}, \qquad
[\wh n, a] = -a.
\ee
The effective representation in chord space corresponds to  the double-scaling limit, \ie the relation 
$\llangle\Tr f(H_{\rm SYK})\rrangle |_{\rm DS}= \mmm{0}{f(H)}{0}$ holds.
The 0-chord state admits a holographic interpretation which identifies it with the infinite temperature thermofield double  state in the dual 
gravitational theory \cite{Lin:2022rbf}.
In \cite{Heller:2024ldz}, it was shown that the finite-temperature Krylov complexity can be written as 
\ba
C(t)_{\beta} &= \frac{1}{Z(\beta)}\sum_{n=0}^{\infty}n\,|\mmm{n}{e^{-iH(t-i\beta/2)}}{0}|^{2} 
= \frac{1}{Z(\beta)}\mmm{0}{e^{iH(t+i\beta/2)}\,\wh n\,e^{-iH(t-i\beta/2)}}{0},
\ea
where $Z(\beta)$ is the Euclidean partition function 
\be
Z(\beta) = \mmm{0}{e^{-\beta H}}{0}.
\ee
Thus, at initial time $t=0$, the quantity $L_{0}(\beta)$ is given by 
\be
\la{2.8}
L_{0}(\beta) = \l\,\frac{\mmm{0}{e^{-\frac{1}{2}\beta H}\wh n e^{-\frac{1}{2}\beta H}}{0}}{\mmm{0}{e^{-\beta H}}{0}}.
\ee

\section{Spectral representation and $L_{0}$ from the two-point function}
\la{sec:spectral}

In the chord basis, the DSSYK Hamiltonian has a continuous spectrum \cite{Berkooz:2018jqr,Berkooz:2018qkz} with energy 
\be
\la{3.1}
E(\theta) = \frac{2J}{\sqrt{\l(1-q)}}\, \cos\theta, \qquad \theta\in[0,\pi],
\ee
and normalized eigenstates $\ket{E(\theta)}$ with chord space wavefunction 
\be
\la{3.2}
\braket{n}{E(\theta)} =  \sqrt{(q;q)_{\infty}}\, |(e^{2i\theta}; q)_{\infty}|\, \frac{H_{n}(\cos\theta|q)}{\sqrt{2\pi(q;q)_{n}}},
\ee
where $H_{n}$ denotes the continuous $q$-Hermite polynomial 
\be
\la{3.3}
H_{n}(\cos\theta|q) \equiv \sum_{k=0}^{n}\frac{(q;q)_{n}}{(q;q)_{n-k}(q;q)_{k}}e^{i(n-2k)\theta}.
\ee
By introducing a complete set of eigenstates of the Hamiltonian $H$, one finds that the numerator and denominator of (\ref{2.8}) have the spectral representation 
\footnote{We use the symbol $\theta$ for the spectral variable of integration, as is customary in the literature on the DSSYK model.
Confusion with the temperature-related parameter in (\ref{1.8}) should not arise.}
\be
\la{3.4}
\mmm{0}{e^{-\frac{1}{2}\beta H}\wh n e^{-\frac{1}{2}\beta H}}{0} = \sum_{n}n\,\bigg[\int_{0}^{\pi}\frac{d\theta}{2\pi}e^{-\frac{\beta}{2}\frac{2J\cos\theta}{\sqrt{\l(1-q)}}}
\frac{(q;q)_{\infty}}{\sqrt{(q;q)_{n}}}|(e^{2i\theta};q)_{\infty}|^{2}H_{n}(\cos\theta|q)
\bigg]^{2},
\ee
\ba
\la{3.5}
Z(\beta)=\mmm{0}{e^{-\beta H}}{0} = (q;q)_{\infty}\int_{0}^{\pi}\frac{d\theta}{2\pi} e^{-\beta\frac{2J\cos\theta}{\sqrt{\l(1-q)}}}|(e^{2i\theta}; q)_{\infty}|^{2}.
\ea
Let us introduce the notation
\be
\la{3.6}
x \equiv x(\beta) = \frac{2J\beta}{\sqrt{\l(1-q)}}.
\ee
The unnormalized two-point function for a matter operator with dimension $\Delta$ \footnote{
Matter operators $\mc O_{\Delta}$ with $0<\Delta<1$ in DSSYK are composites of $p\Delta$  fermions before the 
double-scaling limit, coupled by  random
Gaussian couplings independent of those in the DSSYK Hamiltonian \cite{Berkooz:2018jqr}.}
has the spectral representation \cite{Berkooz:2018jqr}
\be
\la{3.7}
G(x) = \int_{0}^{\pi}\prod_{j=1,2}\frac{d\theta_{j}}{2\pi}e^{-x\cos\theta_{j}}(q;q)_{\infty}|(e^{2i\theta_{j}};q)_{\infty}|^{2}\times \frac{(q^{2\Delta};q)_{\infty}}
{|(q^{\Delta}e^{i(\theta_{1}+\theta_{2})};q)_{\infty}\ (q^{\Delta}e^{i(\theta_{1}-\theta_{2})};q)_{\infty}|^{2}}.
\ee
The normalized 2-point function is obtained by dividing by the partition function \footnote{Here and in the following, with the usual minor abuse of notation, 
we identify $Z(x(\beta))$ with $Z(\beta)$ in (\ref{3.5}), and similarly for
other functions.}
\be
\la{3.8}
\mathscr G(x) = \frac{G(x/2)}{Z(x)},
\ee
which corresponds to insertion points $\beta_{1}=\beta_{2}=\beta/2$.
Relation (\ref{1.6}) becomes
\be
\la{3.9}
L_{0} = -\partial_{\Delta}\mathscr G|_{\Delta=0}.
\ee
Note that (\ref{3.9}) follows immediately from (\ref{2.8}) and the representations (\ref{3.4}) and (\ref{3.5}). To see this, 
we write the $n$ factor in (\ref{3.4}) as $n = -\frac{1}{\l}\partial_{\Delta}q^{n\Delta}$ and 
use the $q$-Mehler formula 
\be
\la{3.10}
\sum_{n=0}^{\infty}\frac{H_{n}(\cos\theta_{1}|q)H_{n}(\cos\theta_{2}|q)}{(q;q)_{n}}\,t^{n} = \frac{(t^{2};q)_{\infty}}{|(te^{i(\theta_{1}+\theta_{2})};q)_{\infty}
(te^{i(\theta_{1}-\theta_{2})};q)_{\infty}|^{2}},
\ee
evaluated at $t=q^{\Delta}$. This gives
\ba
\la{3.11}
& \mmm{0}{e^{-\frac{1}{2}\beta H}\wh n e^{-\frac{1}{2}\beta H}}{0}= 
-\frac{1}{\l}\lim_{\Delta\to 0}\partial_{\Delta}\sum_{n}q^{n\Delta}\,\bigg[\int_{0}^{\pi}\frac{d\theta}{2\pi}e^{-\frac{\beta}{2}\frac{2J\cos\theta}{\sqrt{\l(1-q)}}}
\frac{(q;q)_{\infty}}{\sqrt{(q;q)_{n}}}|(e^{2i\theta};q)_{\infty}|^{2}H_{n}(\cos\theta|q)
\bigg]^{2}\lp
= -\frac{1}{\l}\lim_{\Delta\to 0}\partial_{\Delta}\int_{0}^{\pi}\prod_{j=1,2}\frac{d\theta_{j}}{2\pi}\int_{0}^{\pi}e^{-\frac{\beta}{2}\frac{2J\cos\theta_{j}}{\sqrt{\l(1-q)}}}
(q;q)_{\infty}|(e^{2i\theta_{j}};q)_{\infty}|^{2} \times \frac{(q^{2\Delta};q)_{\infty}}{|(q^{\Delta}e^{i(\theta_{1}+\theta_{2})};q)_{\infty}
(q^{\Delta}e^{i(\theta_{1}-\theta_{2})};q)_{\infty}|^{2}} \lp
= -\frac{1}{\l}\lim_{\Delta\to 0}\partial_{\Delta}G(x/2).
\ea
After dividing by the partition function, \cf (\ref{3.8}), and multiplying by $\l$, we recover  (\ref{3.9}).

\section{Semiclassical expansion of $L_{0}$ in small $\beta$ regime}
\la{sec:small-beta}

We now reproduce the  semiclassical expansion of $L_{0}$  for small $\beta$, Eq.~(\ref{1.17}), in two independent ways: 
from the Krylov definition (\ref{2.8}) using $q$-algebra, and from relation (\ref{3.9}) based on the two-point function.

\subsection{High-temperature expansion of $L_{0}$ from $q$-algebra}

The small-$\beta$ expansion of (\ref{2.8}) is obtained by expanding the exponentials and evaluating the resulting $q$-algebra using (\ref{2.5}) and 
\be
a^{\dagger}\ket{n} = [n+1]_{q}^{1/2}\ket{n+1}, \qquad a\ket{n} = [n]_{q}^{1/2}\ket{n-1},\qquad 
[n]_{q} = \frac{1-q^{n}}{1-q}.
\ee
This gives
\ba
\la{4.2}
& \frac{\mmm{0}{e^{-\frac{1}{2}\beta H}\wh n e^{-\frac{1}{2}\beta H}}{0}}{\mmm{0}{e^{-\beta H}}{0}} =
\frac{1}{4}\bigg(\frac{J\beta}{\sql}\bigg)^{2}+\frac{5 (q-1)}{96}\bigg(\frac{J\beta}{\sql}\bigg)^{4}
+\frac{(q-1)^2 (155+33 q)}{11520}\bigg(\frac{J\beta}{\sql}\bigg)^{6}\lp
+\frac{(q-1)^3 (9331+4872 q+1267 q^2+186 q^3)}{2580480}\bigg(\frac{J\beta}{\sql}\bigg)^{8} +\lp
+\frac{(q-1)^4 (911694 + 790740 q + 399054 q^2 + 150201 q^3 + 41955 q^4 + 8559 q^5 + 
 965 q^6)}{928972800}\bigg(\frac{J\beta}{\sql}\bigg)^{10} +\cdots .
\ea
With the choice $J=1/2$ \footnote{As we recalled in the introduction, the choice $J=1/2$ is the standard convention in the DSSYK/sine-dilaton correspondence.},
we write $\beta$ in terms of $u$, \cf (\ref{1.13}), and, expanding in small $\l$, we  get 
\ba
\la{4.3}
& \l  \frac{\mmm{0}{e^{-\frac{1}{2}\beta H}\wh n e^{-\frac{1}{2}\beta H}}{0}}{\mmm{0}{e^{-\beta H}}{0}} =
u^{2}+\frac{u^{4}}{6}+\frac{2u^{6}}{45}+\frac{17u^{8}}{1260}+\frac{62u^{10}}{14175}+\cdots\lp
+\bigg(\frac{5 u ^4}{12}-\frac{71 u ^6}{180}+\frac{103 u ^8}{252}-\frac{5008u^{10}}{14175}+\cdots\bigg)\l 
+\bigg(-\frac{5 u ^4}{36}+\frac{131 u ^6}{216}-\frac{41549 u 
^8}{30240}+\frac{1607107u^{10}}{680400}+\cdots\bigg)\,\l^{2}\lp
+\bigg(
\frac{5u^{4}}{144}-\frac{101u^{6}}{240}+\frac{5639u^{8}}{2880}-\frac{2312143u^{10}}{388800}+\cdots
\bigg)\, \l^{3}+\cdots.
\ea
This agrees with (\ref{1.17}) and provides the small $u$ expansion of any higher-loop correction in $\l$ at this order in $u$. 
However, extending the expansion to higher powers of u is computationally demanding  by this method, since the $q$-algebra manipulations quickly become intractable.
We note that (\ref{4.2}) is exact in $\l$ and thus one may write each term in the $u$ expansion in a form which is valid at all loop orders, \eg
\be
\la{4.4}
L_{0} = u^{2}+\bigg[1+\frac{5(q-1)}{6\l}\bigg]\,u^{4}+\bigg[\frac{2}{3}+\frac{5(q-1)}{3\l}+\frac{47(q-1)^{2}}{45\l^{2}}+\frac{11(q-1)^{3}}{60\l^{2}}\bigg]\,u^{6}+\cdots.
\ee
As an example, since $q=e^{-\l}$,  the $u^{4}$ bracket resums in closed form to all loop 
orders,
\be
\la{4.5}
\text{coefficient of } u^{4} \text{ in }\,\mc A_{k} = \frac{5(-1)^{k}}{6(k+1)!}, \qquad k=1,2,\ldots,
\ee
reproducing the $u^{4}$ terms in (\ref{4.3}) at $k=1,2,3$ and, for 
instance, predicting $1/144$ at four-loop order.

\subsection{High-temperature expansion of $L_{0}$ from 2-point function}

The partition function $Z(x)$ admits the representation \cite{Berkooz:2018jqr,Xu:2024gfm}
\ba
\la{4.6}
Z(x) &= \sum_{m=-\infty}^{\infty}(-1)^{m}q^{\binom{m}{2}}I_{2m}(x) = \frac{2}{x}\sum_{n=0}^{\infty}(-1)^{n}(2n+1)I_{2n+1}(x)\,q^{n(n+1)/2},
\ea
where $I_{n}(x)$ is the modified Bessel function of the first kind. 
Using 
\be
I_{2n+1}(x) = \sum_{k=0}^{\infty}\frac{1}{k!(k+2n+1)!}(x/2)^{2k+2n+1},
\ee
we get the explicit expansion of $Z(x)$ in small $x$ \footnote{See \cite{Beccaria:2026ndg} for a detailed discussion of the  large $x$ expansion of the partition function.}
\ba
\la{4.8}
Z(x) &= \sum_{p=0}^{\infty}(x/2)^{2p}\sum_{n=0}^{p}(-1)^{n}\frac{2n+1}{(p-n)!(p+n+1)!}\,q^{n(n+1)/2}.
\ea
From Eq.~(C.5) of \cite{Berkooz:2018jqr}, we also have 
\be
\la{4.9}
G(x) = \sum_{n=0}^{\infty}\frac{q^{n\Delta}}{(q;q)_{n}}\sum_{p_{1},p_{2}=0}^{\infty}\prod_{j=1,2}
\frac{2(-1)^{p_{j}}}{x}q^{p_{j}+\binom{p_{j}}{2}}\frac{(q;q)_{n+p_{j}}}{(q;q)_{p_{j}}}(n+2p_{j}+1)\,I_{n+2p_{j}+1}(x).
\ee
The explicit expansion at small $x$ is 
\ba
G(x) &= 1-\frac{q-1}{4}x^{2}+\frac{(q-1)^{2}(5+q)}{192}x^{4}-\frac{(q-1)^{3}(35+21q+3q^{2}+q^{3})}{23040}x^{6}+\cdots\lp
+\bigg[-\frac{q-1}{4}x^{2}+\frac{(q-1)^{2}(2+q)}{48}x^{4}-\frac{(q-1)^{3}(35+38q+14q^{2}+3q^{3})}{11520}x^{6}+\cdots\bigg]q^{\Delta}\lp
+\bigg[\frac{(q-1)^{2}(1+q)}{64}x^{4}-\frac{(q-1)^{3}(q+1)(3+2q+q^{2})}{1536}x^{6}+\cdots \bigg]q^{2\Delta}+\cdots.
\ea
Using (\ref{4.8}), one has
\be
Z(x) = 1-\frac{q-1}{8}x^{2}+\frac{(q-1)^{2}(2+q)}{384}x^{4}-\frac{(q-1)^{3}(5+6q+3q^{2}+q^{3})}{46080}x^{6}+\cdots.
\ee
The normalized 2-point function is given by (\ref{3.8}), where we recall that 
the relation between $x$ and $\beta, \l$ is as in (\ref{3.6}).
An important  observation is that each power $x^{n}$ receives contributions from a finite number of $q^{p\Delta}$ terms. The computation of 
$L_{0}$ from (\ref{3.9}) thus requires the analysis of a finite expression and one gets
\be
\frac{1}{\l}L_{0} = -\frac{1}{\l}\partial_{\Delta}\mathscr G|_{\Delta=0} = \frac{1}{4}\bigg(\frac{J\beta}{\sql}\bigg)^{2}+\frac{5(-1+q)}{96}\bigg(\frac{J\beta}{\sql}\bigg)^{4}
+\frac{(-1+q)^{2}(155+33q)}{11520}\bigg(\frac{J\beta}{\sql}\bigg)^{6}+\cdots,
\ee
in agreement with (\ref{4.2}).

\section{Two-loop saddle point expansion}
\la{sec:saddle}

The spectral representation of the two-point function, together with the relation (\ref{3.9}), allows one to compute analytically 
the one-loop expansion (\ref{1.16}), first obtained in \cite{Goel:2023svz,Okuyama:2023bch}. In this section, we extend the calculation to the two-loop level. 
Our goal is to show that the special structure of the
one-loop correction in (\ref{1.16}) is generic, \ie we expect to find a linear combination of $\tan^{2n} u$ and $u\tan^{2n+1}u$ terms with coefficients that are polynomials in $u^{2}$ and 
an overall factor which is an integer inverse power of $1+u\tan u$. A similar uniform structure in higher loop corrections 
to the quantum length at $\beta=0$ and generic time is discussed in the 5-loop computation in \cite{Alfinito:2026cky}.

Our starting point is (\ref{3.7}), rewritten in terms of the insertion points $\beta_{i}$ \footnote{Here we absorb $J$ into the definition of  $\beta_{i}$ by $J\beta_{i}\to \beta_{i}$. }
\be
\la{5.1}
G = \int_{0}^{\pi}\prod_{j=1,2}\frac{d\theta_{j}}{2\pi}e^{2\frac{\beta_{j}}{\sqrt{\l(1-q)}}\cos\theta_{j}}(q;q)_{\infty}|(e^{2i\theta_{j}};q)_{\infty}|^{2}\  \frac{(q^{2\Delta};q)_{\infty}}
{|(q^{\Delta}e^{i(\theta_{1}+\theta_{2})};q)_{\infty}\ (q^{\Delta}e^{i(\theta_{1}-\theta_{2})};q)_{\infty}|^{2}}\ .
\ee
We need this for $q\to 1$ or, equivalently, $\l\to 0^{+}$. In this section, we review the one-loop computation in \cite{Goel:2023svz,Okuyama:2023bch} in full detail
to set up notation and to clarify  the  cancellation of terms which are singular in the $\Delta\to 0$ limit, which is a crucial fact to apply (\ref{3.9}).

In the following, we shall repeatedly use the formula \cite{Okuyama:2023bch}
\be
\la{5.2}
(a;q)_{\infty} = \exp\bigg[-\sum_{g=0}^{\infty}\frac{\l^{2g-1}B_{2g}}{(2g)!}\LI_{2-2g}(a)+\frac{1}{2}\log(1-a)\bigg],
\ee
which applies to constant $a$. This gives, for the spectral measure terms, the standard expression
\be
\la{5.3}
(q;q)_{\infty}|(e^{2i\theta}; q)_{\infty}|^{2} = \sqrt\frac{2\pi}{\l}\exp\bigg[\frac{\l}{8}-\frac{2}{\l}\bigg(\theta-\frac{\pi}{2}\bigg)^{2}+\log(2\sin\theta)\bigg]+\text{n.p.},
\ee
where ``n.p.'' stands for higher non-perturbative corrections that we omit since we focus on the perturbative part of the semiclassical expansion.

Before turning to the details, it is useful to outline the strategy. We first identify the 
classical saddle point of the $\theta_{1},\theta_{2}$ integral in (\ref{5.1}), which at the 
symmetric point $\phi=0$ relevant for $L_{0}$ has the two angles coincide. 
Section~\ref{sec:one-loop} then expands around this saddle to one loop: the fluctuation 
determinant develops individual poles at the coincident point, which cancel only once assembled 
with the classical prefactor, and only if the regulator $\eps\equiv\l\Delta$ is treated as 
independent of $\l$ throughout the calculation, rather than set to zero from the start. 
Section~\ref{two-loop} repeats this one order further in the fluctuation expansion, where the 
same cancellation mechanism must now resolve individual poles of order up to $1/\eps^{3}$.

\subsection{Classical saddle point}

At leading order, we write (\ref{5.1}) as 
\be
G = \frac{2\pi}{\l}\int_{0}^{\pi}\prod_{j=1,2}\frac{d\theta_{j}}{2\pi}e^{-F},
\ee
with 
\ba
\la{5.5}
F &= -2\bigg(\frac{1}{\l}+\frac{1}{4}+O(\l)\bigg)(\beta_{1}\cos\theta_{1}+\beta_{2}\cos\theta_{2})+\frac{2}{\l}\bigg(\theta_{1}-\frac{\pi}{2}\bigg)^{2}+\frac{2}{\l}\bigg(\theta_{2}-\frac{\pi}{2}\bigg)^{2}\lp
-\log(2\sin\theta_{1})-\log(2\sin\theta_{2}) \lp
-\frac{1}{\l}\LI_{2}(q^{\Delta}e^{i(\pm\theta_{1}\pm\theta_{2})})
+\frac{1}{2}\log(1-q^{\Delta}e^{i(\pm\theta_{1}\pm\theta_{2})})-\log(q^{2\Delta}; q)+O(\l),
\ea
where in each term with $\pm$  we sum over all 4 possible sign combinations. The last term in (\ref{5.5}) is from the numerator of the last factor in (\ref{5.1}).
Let us write
\be
\la{5.6}
q^{\Delta} = e^{-\eps}, \qquad \eps>0,
\ee
and introduce the notation -- useful for later discussion --
\be
\Lambda_{1}(\eps,\alpha) = \LI_{2}(e^{-\eps}e^{i\alpha})+\LI_{2}(e^{-\eps}e^{-i\alpha}), \qquad \Lambda_{2}(\eps,\alpha) = \log(1-e^{-\eps}e^{i\alpha})+\log(1-e^{-\eps}e^{-i\alpha}).
\ee
The relation (\ref{5.6}) implies $\eps = \l\Delta$. However, as explained in \cite{Goel:2023svz}, the correct way to deal with the saddle point expansion is to consider
$\eps$ as a small quantity independent of $\l$.

Accordingly,  we use (\ref{5.2}) to get 
\be
-\log(q^{2\Delta}; q)_{\infty} = -\log(e^{-2\eps}; q)_{\infty} = \frac{1}{\l}\LI_{2}(e^{-2\eps})-\frac{1}{2}\log(1-e^{-2\eps})+O(\l),
\ee
and write
\ba
\la{5.9}
F &= -2\bigg(\frac{1}{\l}+\frac{1}{4}+O(\l)\bigg)(\beta_{1}\cos\theta_{1}+\beta_{2}\cos\theta_{2})+\frac{2}{\l}\bigg(\theta_{1}-\frac{\pi}{2}\bigg)^{2}+\frac{2}{\l}\bigg(\theta_{2}-\frac{\pi}{2}\bigg)^{2}\lp
-\log(2\sin\theta_{1})-\log(2\sin\theta_{2}) \lp
-\frac{1}{\l}\Lambda_{1}(\eps,\theta_{1}+\theta_{2})-\frac{1}{\l}\Lambda_{1}(\eps,\theta_{1}-\theta_{2})
+\frac{1}{2}\Lambda_{2}(\eps,\theta_{1}+\theta_{2})+\frac{1}{2}\Lambda_{2}(\eps,\theta_{1}-\theta_{2})\lp
+\frac{1}{\l}\LI_{2}(e^{-2\eps})-\frac{1}{2}\log(1-e^{-2\eps})+O(\l).
\ea
After writing
\be
\la{5.10}
F = \frac{1}{\l}F_{0}+F_{1}+\l F_{2}+\cdots, 
\ee
we have thus the leading contribution 
\ba
\la{5.11}
F_{0} &= -2(\beta_{1}\cos\theta_{1}+\beta_{2}\cos\theta_{2})+2\bigg(\theta_{1}-\frac{\pi}{2}\bigg)^{2}+2\bigg(\theta_{2}-\frac{\pi}{2}\bigg)^{2}
-\Lambda_{1}(\eps,\theta_{1}+\theta_{2})-\Lambda_{1}(\eps,\theta_{1}-\theta_{2})+\LI_{2}(e^{-2\eps}).
\ea
We now use the relations
\bea
\Lambda_{1}(\eps,\alpha) &=  \sum_{n=1}^{\infty}\frac{2}{n^{2}}e^{-n\eps}\cos(n\alpha), \\
\frac{d}{d\alpha}\Lambda_{1}(\eps, \alpha) &= -\sum_{n=1}^{\infty}\frac{2}{n}e^{-n\eps}\sin(n\alpha) = -2\arctan\frac{\sin\alpha}{e^{\eps}-\cos\alpha},
\eea
and get the leading order saddle equations as 
\bea
\la{5.13}
& \beta_{1}\sin\theta_{1}+2\theta_{1}-\pi+\arctan\frac{\sin(\theta_{1}+\theta_{2})}{e^{\eps}-\cos(\theta_{1}+\theta_{2})}
+\arctan\frac{\sin(\theta_{1}-\theta_{2})}{e^{\eps}-\cos(\theta_{1}-\theta_{2})}=0, \\
& \beta_{2}\sin\theta_{2}+2\theta_{2}-\pi+\arctan\frac{\sin(\theta_{1}+\theta_{2})}{e^{\eps}-\cos(\theta_{1}+\theta_{2})}
+\arctan\frac{\sin(\theta_{2}-\theta_{1})}{e^{\eps}-\cos(\theta_{2}-\theta_{1})}=0.
\eea
The Ansatz for the solution is \cite{Goel:2023svz,Okuyama:2023bch}
\be
\la{5.14}
\theta_{i} = \frac{\pi}{2}-u+a_{i}\,\eps+b_{i}\,\eps^{2}+O(\eps^{3}).
\ee
Substituting this in (\ref{5.13}) with 
\be
\beta_{1}=\frac{u-\phi}{\cos u}, \qquad \beta_{2}=\frac{u+\phi}{\cos u},
\ee
and expanding in $\eps$ gives at order $O(\eps^{0})$
\be
\la{5.16}
\beta\equiv \beta_{1}+\beta_{2}=\frac{2u}{\cos u}, \qquad a_{1}-a_{2} = \tan\phi.
\ee
Going at $O(\eps)$ provides the sum 
\be
a_{1}+a_{2} = \frac{\tan u(1+\phi\tan\phi)}{1+u\tan u},
\ee
and thus separately $a_{1}, a_{2}$. We also get the difference 
\be
b_{1}-b_{2} = \frac{1}{2\cos^{2}\phi}\bigg[-(1+u\tan u)\tan\phi+\frac{\phi\tan^{2}u(1+\phi\tan\phi)}{1+u\tan u}\bigg].
\ee
The order $O(\eps^{2})$ determines $b_{1}+b_{2}$ (which we do not write explicitly) and so on.

Now, we need to substitute (\ref{5.14}) into (\ref{5.11}) and expand in $\eps\to 0^{+}$. 
Since we are interested in $L_{0}$ that corresponds to insertions at $\beta_{1}=\beta_{2}=\beta/2$, we specialize to $\phi=0$. 
A short computation using the relations ($0<u<\pi/2$)
\bea
 & \LI_{2}(-e^{-2iu})+\LI_{2}(-e^{2iu}) = -\frac{\pi^{2}}{6}+2u^{2}, \qquad
\log(1+e^{-2iu})+\log(1+e^{2iu}) = \log(4\cos^{2}u), \\
& \log(1+e^{-2iu})-\log(1+e^{2iu}) = -2iu,
\eea
gives
\be
\la{5.20}
F_{0} = u(2u-4\tan u)-\log(\cos^{2}u)\,\eps+O(\eps^{2}).
\ee
Thus, the classical approximation of the  normalized 2-point function at $\phi=0$ reads
\be
\mathscr G = e^{-\frac{1}{\l}(F_{0}-F_{0}|_{\eps=0})} = (\cos^{2}u)^{\eps/\l} =  (\cos^{2}u)^{\Delta}.
\ee

\subsection{Saddle expansion at one-loop order}
\la{sec:one-loop}

We organize the small-$\l$ expansion of $F$ as in (\ref{5.10})
with $F_{0}$ as in (\ref{5.11})--(\ref{5.20}) and, reading off the $O(\l^{0})$ terms of (\ref{5.9}),
\ba
\la{5.22}
F_{1} &= -\frac{1}{2}(\beta_{1}\cos\theta_{1}+\beta_{2}\cos\theta_{2})-\log(2\sin\theta_{1})-\log(2\sin\theta_{2})\lp
+\frac{1}{2}\Lambda_{2}(\eps,\theta_{1}+\theta_{2})+\frac{1}{2}\Lambda_{2}(\eps,\theta_{1}-\theta_{2})-\frac{1}{2}\log(1-e^{-2\eps}).
\ea
We now  include the Gaussian fluctuations around the saddle by setting
\be
\la{5.23}
\theta_{i} = \theta_{i}^{*}+\sqrt\l\,\xi_{i}, \qquad i=1,2,
\ee
where $\theta_{i}^{*}$ solves (\ref{5.13}), \ie $\nabla F_{0}(\theta^{*})=0$. Expanding (\ref{5.10}) to quadratic order in the 
fluctuations,
\be
\la{5.24}
F(\theta^{*}+\sqrt\l\,\xi) = \frac{1}{\l}F_{0}(\theta^{*})+\frac{1}{2}\,\xi^{T}H\xi+F_{1}(\theta^{*})+O(\sqrt\l),
\ee
where $H_{ij}=\partial^{2}F_{0}/\partial\theta_{i}\partial\theta_{j}\big|_{\theta^{*}}$ is the Hessian of $F_{0}$ at the saddle,
\bea
\la{5.25}
H_{11} &= 2\beta_{1}\cos\theta_{1}+4-\Lambda_{1}''(\eps,\theta_{1}+\theta_{2})-\Lambda_{1}''(\eps,\theta_{1}-\theta_{2}), \\
H_{22} &= 2\beta_{2}\cos\theta_{2}+4-\Lambda_{1}''(\eps,\theta_{1}+\theta_{2})-\Lambda_{1}''(\eps,\theta_{1}-\theta_{2}), \\
H_{12} &= -\Lambda_{1}''(\eps,\theta_{1}+\theta_{2})+\Lambda_{1}''(\eps,\theta_{1}-\theta_{2}),
\eea
with 
\be
\la{5.26}
\Lambda_{1}''(\eps,\alpha)\equiv\partial_{\alpha}^{2}\Lambda_{1}(\eps,\alpha) = -2\sum_{n=1}^{\infty}e^{-n\eps}\cos(n\alpha) 
= -2\,\dfrac{e^{-\eps}\cos\alpha-e^{-2\eps}}{1-2e^{-\eps}\cos\alpha+e^{-2\eps}}. 
\ee
Doing the Gaussian integral over $\xi_{1},\xi_{2}$ 
and collecting the prefactors in (\ref{5.23}) gives the one-loop approximation
\be
\la{5.27}
G \approx e^{-F_{0}(\theta^{*})/\l}\,B(u,\phi,\eps), \qquad B(u,\phi,\eps)\equiv \frac{e^{-F_{1}(\theta^{*})}}{\sqrt{\det H}}.
\ee
The partition function $Z$ is independent of $\Delta$ and the normalized two-point function 
$\sG \equiv G/Z$ must equal $1$ at $\eps=0$. Thus,  we do not need to compute $Z$. We simply get 
\be
\la{5.28}
\sG = (\cos^{2}u)^{\Delta}\bigg[1+\eps\,\mc A_{1}(u,\phi)+O(\eps^{2})\bigg], \qquad 
 \mc A_{1}(u,\phi) \equiv \frac{d}{d\eps}\log\frac{B(u,\phi,\eps)}{B(u,\phi,0)}\bigg|_{\eps=0}.
\ee
Note that this is consistent with  (\ref{1.9}) and (\ref{1.10})   since  $\eps = \Delta\l$.

\paragraph{Specialized evaluation at $\phi=0$.}
We remark that at $\phi=0$ we have $\theta_{1}-\theta_{2}=0$ through $O(\eps^{2})$, and both 
$\Lambda_{1}''(\eps,\theta_{1}-\theta_{2})$ and $\Lambda_{2}(\eps,\theta_{1}-\theta_{2})$ are then evaluated at the coincident 
point $\theta_{1}=\theta_{2}$, where they individually diverge as $\eps\to0$:
\ba
\la{5.29}
\Lambda_{1}''(\eps,0) &= -\frac{2e^{-\eps}}{1-e^{-\eps}} = -\frac{2}{\eps}+1-\frac{\eps}{6}+O(\eps^{2}), \\
\la{5.30}
\frac{1}{2}\Lambda_{2}(\eps,0)-\frac{1}{2}\log(1-e^{-2\eps}) &= \frac{1}{2}\log\tanh\frac{\eps}{2} = \frac{1}{2}\log\eps-\frac{1}{2}\log2+O(\eps^{2}).
\ea
So individually $\det H\sim 1/\eps$ and $F_{1}\sim \frac12\log\eps$ as $\eps\to0$ at $\phi=0$. We will see below that 
these cancel exactly in $B(u,0,\eps)$, which has a finite limit as $\eps\to0$.

At $\phi=0$ we have $\beta_{1}=\beta_{2}=u/\cos u$ and $\theta_{1}=\theta_{2}=\pi/2-u+a\eps+O(\eps^{2})$ with 
$a=\tan u/(2(1+u\tan u))$ from (\ref{5.14})--(\ref{5.13}). We expand $H_{ij}$ and $F_{1}$ to $O(\eps)$, splitting each 
$\Lambda_{1}'',\Lambda_{2}$ into the regular ``$+$'' direction $\alpha=\theta_{1}+\theta_{2}=\alpha_{0}+2a\eps$, 
$\alpha_{0}=\pi-2u$, and the singular ``$-$'' direction $\theta_{1}-\theta_{2}=0$. For the regular direction, ordinary 
Taylor expansion in $\eps$ gives
\be
\la{5.31}
\Lambda_{1}''(0,\alpha_{0}) = 1, \qquad 
P\equiv\partial_{\eps}\Lambda_{1}''(\eps,\alpha_{0})\big|_{\eps=0} = -\frac{1}{2\cos^{2}u}, \qquad 
Q\equiv\partial_{\alpha}\Lambda_{1}''(0,\alpha)\big|_{\alpha_{0}} = 0,
\ee
\be
\la{5.32}
\Lambda_{2}(0,\alpha_{0}) = \log(4\cos^{2}u), \qquad
S\equiv\partial_{\eps}\Lambda_{2}(\eps,\alpha_{0})\big|_{\eps=0} = -1, \qquad
T\equiv\partial_{\alpha}\Lambda_{2}(0,\alpha)\big|_{\alpha_{0}} = \tan u,
\ee
while for the singular direction we use (\ref{5.29}), whose $O(\eps)$ coefficient we denote $R\equiv-1/6$ (there is no $O(\eps)$ 
term in  (\ref{5.30})). Assembling (\ref{5.25}) along $\theta_{1}=\theta_{2}=\pi/2-u+a\eps$, we get
\be
\la{5.33}
H_{11}=H_{22} = \frac{2}{\eps}+C_{1}+\eps\,D_{1}+O(\eps^{2}), \qquad H_{12} = -\frac{2}{\eps}+\eps\,E+O(\eps^{2}),
\ee
with
\be
\la{5.34}
C_{1} = 2(1+u\tan u), \qquad D_{1} = -2au-P-R, \qquad E = R-P.
\ee
Then $\det H = H_{11}H_{22}-H_{12}^{2} = \dfrac{4C_{1}}{\eps}+K+O(\eps)$, with $K\equiv C_{1}^{2}+4D_{1}+4E$, so
\be
\la{5.35}
\frac{1}{\sqrt{\det H}} = \frac{1}{2}\sqrt{\frac{\eps}{C_{1}}}\bigg[1-\frac{K}{8C_{1}}\,\eps+O(\eps^{2})\bigg].
\ee
Similarly, assembling $F_{1}$ in (\ref{5.22}) using (\ref{5.29})--(\ref{5.32}),
\be
\la{5.36}
F_{1} = -\log(2\cos u)-u\tan u+\frac{1}{2}\log\eps-\frac{1}{2}\log2+F_{1}^{(1)}\,\eps+O(\eps^{2}), \qquad F_{1}^{(1)} = a(u-\tan u)-\frac{1}{2},
\ee
so that
\be
\la{5.37}
e^{-F_{1}} = 2\sqrt2\,\cos u\,e^{u\tan u}\,\eps^{-1/2}\Big[1-F_{1}^{(1)}\eps+O(\eps^{2})\Big].
\ee
The $\eps^{\mp1/2}$ factors between (\ref{5.35}) and (\ref{5.37}) cancel exactly, as anticipated above, and give
\be
\la{5.38}
B(u,0,\eps) = \frac{\cos u\,e^{u\tan u}}{\sqrt{1+u\tan u}}\bigg[1+D(u)\,\eps+O(\eps^{2})\bigg], \qquad D(u) \equiv -\bigg(F_{1}^{(1)}+\frac{K}{8C_{1}}\bigg).
\ee
Comparing with (\ref{5.28}) gives $\mc A_{1}(u,0)=D(u)$. Substituting 
$a=\tan u/(2(1+u\tan u))$ and (\ref{5.31})--(\ref{5.34}) and simplifying,
\be
\la{5.39}
\mc A_{1}(u,0) = \frac{-u\tan u+(1-3u^{2})\,\tan^{2}u+u(1-u^{2})\, \tan^{3}u}{4(1+u\tan u)^{2}},
\ee
which is correctly minus the $O(\l)$ term in (\ref{1.16}).
Note that the cancellation of the individually singular pieces in (\ref{5.29}), verified here at $\phi=0$, persists for 
general $\phi$ as well, leading to the finite general expression in (\ref{1.11}).

\subsection{Extension to two-loop order}
\la{two-loop}

Having clarified the procedure at one-loop order, including manifest cancellation of singular terms, we now move on to the novel two-loop
correction. The first piece to analyze is $F_{2}$, the $O(\l)$ term of the fluctuation expansion (\ref{5.10}).
Using the elementary identity (from $\LI_{s-1}(z)=z\,d/dz\,\LI_{s}(z)$ applied twice to $\Lambda_{1}$)
\be
\la{5.40}
\Lambda_{0}(\eps,\alpha) \equiv \LI_{0}(e^{-\eps+i\alpha})+\LI_{0}(e^{-\eps-i\alpha}) = -\Lambda_{1}''(\eps,\alpha),
\ee
the $g=1$ term of (\ref{5.2}) applied to the numerator and denominator factors of (\ref{5.1}), together with the exact $\l/8$ in (\ref{5.3}) and the elementary 
expansion $1/\sqrt{\l(1-q)}=1/\l+1/4+\l/96+O(\l^{2})$, combine into
\be
\la{5.41}
F_{2} = -\frac14 - \frac{1}{48}(\beta_{1}\cos\theta_{1}+\beta_{2}\cos\theta_{2}) + \frac{1}{12}\frac{e^{-2\eps}}{1-e^{-2\eps}} 
+ \frac{1}{12}\Big[\Lambda_{1}''(\eps,\theta_{1}+\theta_{2})+\Lambda_{1}''(\eps,\theta_{1}-\theta_{2})\Big],
\ee
where we recall that the explicit expression for $\Lambda_{1}''$ was given in (\ref{5.26}).

\paragraph{Fluctuations to next order.} Next, we need to study fluctuations and expand (\ref{5.10}) one order further in the perturbation (\ref{5.23}),
\be
\la{5.42}
F(\theta^{*}+\sqrt\l\,\xi) = \frac{F_{0}(\theta^{*})}{\l}+\frac{1}{2}\xi^{T}H\xi + F_{1}(\theta^{*}) + \sqrt\l\,V_{3} + \l\big(V_{4}+F_{2}(\theta^{*})\big) + O(\l^{3/2}),
\ee
where
\be
\la{5.43}
V_{3} = \frac16 F_{0,ijk}\,\xi^{i}\xi^{j}\xi^{k} + F_{1,i}\,\xi^{i}, \qquad V_{4} = \frac{1}{24}F_{0,ijkl}\,\xi^{i}\xi^{j}\xi^{k}\xi^{l} + \frac12 F_{1,ij}\,\xi^{i}\xi^{j}.
\ee
In the Gaussian average, we have $\langle V_{3}\rangle=0$. Nonzero contributions come from the expansion of  $e^{-\sqrt\l V_{3}-\l(V_{4}+F_{2})}$ and averaging. 
This shifts $B$ in  (\ref{5.27}) according to 
\be
\la{5.44}
B(u,\phi,\eps,\l) = B(u,\phi,\eps)\Big[1+\l\,\sC(u,\phi,\eps)+O(\l^{2})\Big], \qquad \sC \equiv \frac12\langle V_{3}^{2}\rangle - \langle V_{4}\rangle - F_{2}(\theta^{*}),
\ee
so that, repeating the argument leading to (\ref{5.28}), we get (for general $\phi$)
\be
\la{5.45}
\mc A_{2}(u,\phi) \equiv \frac{d}{d\eps}\Big[\sC(u,\phi,\eps)-\sC(u,\phi,0)\Big]\bigg|_{\eps=0}.
\ee
We now evaluate this quantity in the simplified case $\phi=0$. Diagonalize the fluctuations as $\eta_{\pm}=(\xi_{1}\pm\xi_{2})/\sqrt2$, so $H_{\pm\pm}\equiv H_{11}\pm H_{12}$ are the corresponding eigenvalues; from (\ref{5.33}), $H_{++}=C_{1}+O(\eps)$ is regular while $H_{--}\sim4/\eps$ is singular, as before. Since $F_{0},F_{1}$ are even under $\theta_{1}\leftrightarrow\theta_{2}$ at $\phi=0$, $V_{3},V_{4}$ contain only even powers of $\eta_{-}$.
We have
\bea
\la{5.46}
V_{3} &= A\,\eta_{+}^{3}+B\,\eta_{+}\eta_{-}^{2}+C\,\eta_{+}, \\
V_{4} &= P_{1}\eta_{+}^{4}+P_{2}\eta_{+}^{2}\eta_{-}^{2}+P_{3}\eta_{-}^{4}+P_{4}\eta_{+}^{2}+P_{5}\eta_{-}^{2},
\eea
built from the derivatives $F_{0,111}=F_{0,222}$, $F_{0,112}=F_{0,122}$, $F_{1,1}=F_{1,2}$ (for $V_{3}$) and $F_{0,1111}=F_{0,2222}$, $F_{0,1112}=F_{0,1222}$, $F_{0,1122}$, $F_{1,11}=F_{1,22}$, $F_{1,12}$ (for $V_{4}$), in the pattern of (\ref{5.25}).

We remark here that 
several ingredients  are singular as $\eps\to0$, more severely than at one loop: $\Lambda_{1}''''(\eps,0)\sim4/\eps^{3}$ and $\Lambda_{2}''(\eps,0)\sim2/\eps^{2}$ 
both appear (in $P_{3}$ and $P_{5}$ respectively), compared to $\Lambda_{1}''(\eps,0)\sim-2/\eps$ in (\ref{5.29}). 
These cancel exactly against a matching singularity supplied by $F_{2}(\theta^{*})$ once (\ref{5.44}) is assembled, 
leaving $\sC(\eps)$ finite as $\eps\to0$.

An exact simplification controls the ``sum''-direction ($\alpha=\theta_{1}+\theta_{2}$) pieces entering (\ref{5.46}). The classical Fourier series
\be
\la{5.47}
\Lambda_{1}(0,\alpha) = \frac{\pi^{2}}{3}-\pi\alpha+\frac{\alpha^{2}}{2}, \qquad 0<\alpha<2\pi,
\ee
is an \emph{exact} quadratic polynomial in $\alpha$. Hence $\Lambda_{1}''(0,\alpha)=1$ (\cf (\ref{5.31})) and, crucially, \emph{every higher $\alpha$-derivative vanishes identically}: $\Lambda_{1}'''(0,\alpha)=\Lambda_{1}''''(0,\alpha)=0$ throughout $\alpha\in(0,2\pi)$, not merely at $\alpha=0$. Their small-$\eps$ behaviour is then governed entirely by the direct $\eps$-derivative,
\be
\la{5.48}
K(\alpha)\equiv\partial_{\eps}\Lambda_{1}'''(\eps,\alpha)\big|_{\eps=0} = \frac{\sin\alpha}{(\cos\alpha-1)^{2}}, \qquad 
M(\alpha)\equiv\partial_{\eps}\Lambda_{1}''''(\eps,\alpha)\big|_{\eps=0} = \frac{\sin^{2}\alpha-\cos\alpha+1}{(\cos\alpha-1)^{3}},
\ee
since the saddle shift $s^{*}=\alpha_{0}+2a\eps$, $\alpha_{0}=\pi-2u$, contributes nothing at $O(\eps)$ ($\Lambda_{1}''''(0,\alpha)=0$ kills the chain-rule term). An analogous, if less complete, simplification holds for $\Lambda_{2}$: $\partial_{\eps}\Lambda_{2}'(\eps,\alpha)|_{0}=\partial_{\eps}\Lambda_{2}''(\eps,\alpha)|_{0}=0$ identically, so $\Lambda_{2}'(\eps,s^{*})$ and $\Lambda_{2}''(\eps,s^{*})$ are likewise fixed at $O(\eps)$ purely by the $\alpha$-shift, via
\be
\la{5.49}
\Lambda_{2}''(0,\alpha)=\frac{1}{\cos\alpha-1}, \qquad \Lambda_{2}'''(0,\alpha) = K(\alpha).
\ee
Assembling (\ref{5.46}) using (\ref{5.47})--(\ref{5.49}) at $\theta_{1}=\theta_{2}=\pi/2-u+a\eps$, $s^{*}=\alpha_{0}+2a\eps$,
we get the coefficients
\bea
\la{5.50}
A &= -\frac{u}{3\sqrt2} - \frac{\eps}{3\sqrt2}\bigg(au\tan u+\frac{\sin u}{\cos^{3}u}\bigg), \qquad
B = -\frac{u}{\sqrt2} - \frac{\eps\,au\tan u}{\sqrt2}, \\
C &= \frac{u-\tan u}{\sqrt2} + \frac{\eps\,a}{\sqrt2}\big(u\tan u+\sec^{2}u\big),
\eea
and also 
\bea
\la{5.51}
P_{1} &= -\frac{u\tan u}{24} + \eps\bigg(\frac{au}{24}-\frac{M(\alpha_{0})}{6}\bigg), \qquad
P_{2} = -\frac{u\tan u}{4} + \frac{\eps\,au}{4}, \\
P_{3} &= -\frac{2}{3\eps^{3}} - \frac{u\tan u}{24} + \eps\bigg(\frac{au}{24}+\frac{1}{360}\bigg), \\
P_{4} &= \frac{u\tan u}{4}+\frac{1}{4\cos^{2}u} + \eps\bigg(-\frac{au}{4}-\frac{a\tan u}{\cos^{2}u}+\frac{a\sin u}{2\cos^{3}u}\bigg), \\
P_{5} &= \frac{1}{\eps^{2}} + \frac{u\tan u}{4}+\frac{1}{2\cos^{2}u}-\frac{1}{12} + \eps\bigg(-\frac{au}{4}-\frac{a\tan u}{\cos^{2}u}\bigg),
\eea
with $M(\alpha_{0})=(2\cos^{2}u-3)/(4\cos^{4}u)$, and
\be
\la{5.52}
F_{2}(\theta^{*}) = -\frac{1}{8\eps} - \frac18 - \frac{u\tan u}{24} + \frac{\eps}{24}\big(au-\sec^{2}u\big) + O(\eps^{2}).
\ee
Together with $H_{\pm\pm}$ from (\ref{5.33}) (and one further order, $H_{--}=4/\eps+C_{1}+\eps(1/3-2au)-\eps^{2}a^{2}u\tan u+
O(\eps^{3})$), equations (\ref{5.44})--(\ref{5.45}) can now be assembled. The $1/\eps^{3},1/\eps^{2},1/\eps$ singularities cancel exactly, leaving the final two-loop result
\ba
\la{5.53}
\mc A_{2}(u,0) &= \frac{1}{192\,(1+u\tan u)^{5}}\,[
-9u^{2}+(13-30u^{2})\,u\,\tan u+(-4+91u^{2}-66u^{4})\,\tan^{2}u \lp
-6(7-26u^{2}+7u^{4})\,u\,\tan^{3}u 
-2(-3+54u^{2}-67u^{4}+6u^{6})\,\tan^{4}u \lp
+2(9-61u^{2}+40u^{4})\,u\,\tan^{5}u
+2u^{2}(9-34u^{2}+15u^{4})\,\tan^{6}u+2u^{3}(3-8u^{2}+3u^{4})\,\tan^{7}u].
\ea
Its expansion at small $u$ (corresponding to high temperature) reads
\be
\mc A_{2}(u,0) = \frac{5 u^4}{36}-\frac{131 u^6}{216}+\frac{41549 
u^8}{30240}-\frac{1607107 u^{10}}{680400}+\frac{153828139 
u^{12}}{44906400}-\frac{9048005971 
u^{14}}{2043241200}+\cdots,
\ee
which reproduces minus the $O(\l^{2})$ terms in (\ref{4.3}), the change of sign being due to the minus in (\ref{3.9}).

\section{An alternative strategy for the three-loop contribution}
\la{sec:3loop}

The explicit two-loop correction (\ref{5.53}) remarkably confirms the structure of the one-loop correction in (\ref{5.39}). In other words, we expect the $k$-loop correction 
to take the form 
\be
\la{6.1}
\mc A_{k}(u,0) = \frac{\sum_{p=0}^{N_{2}}\mc P_{p}(u^{2})\, u^{\sigma_{p}}\tan^{p}u}{(1+u\tan u)^{N_{1}}}, \qquad \sigma_{p} = \begin{cases}
0, & p\ \text{even}, \\
1, & p\ \text{odd},
\end{cases}
\ee
where $N_{1}, N_{2}$ depend on $k$ and $\mc P_{p}(z)$ are (also $k$-dependent) polynomials with maximum degree 1 at one-loop and 3 at two-loop. 
This immediately suggests a practical strategy for determining the three-loop correction without performing the full next-order saddle point calculation.
If we can extend the small $u$ expansion in (\ref{4.3}) and make it 
long enough, we can test the Ansatz (\ref{6.1}) with reasonable values of $N_{1},N_{2}$ and the degree of the polynomials $\mc P_{p}$, using a certain number of terms in the
small $u$ expansion to fix the arbitrary coefficients, and testing them against additional data points. However, this is far from trivial: based on the one-loop and two-loop expressions,
a natural assumption is to have 
\be
\la{6.2}
N_{1}=3k-1, \qquad N_{2}=4k-1, \qquad \deg \mc P_{p}\le 2k-1,
\ee
where $k$ is the loop order. At three-loop order this gives 
$N_{1}=8$, $N_{2}=11$, and the degree of $\mc P_{p}\le 5$. This results in $6\times 12=72$ unknown coefficients and requires the small $u$ expansion (which is actually in $u^{2}$)
to arrive at $u^{144}$ plus other terms to be used for validation. The methods in Section~\ref{sec:small-beta} cannot be used for such a task. 
\footnote{We have verified that no consistent solution exists for the Ansatz (\ref{6.1}) with 
degrees smaller than (\ref{6.2}): the values of $N_{1},N_{2},\deg\mc P_{p}$ are not merely 
sufficient but minimal.}
In this section, we begin by deriving an efficient algorithm for the evaluation of long Taylor series in $u$ of the various loop order 
contributions to $L_{0}$ in the semiclassical limit. This is interesting in its own right, since it will reveal a nice recursion relation hidden in the 
explicit formula (\ref{4.9}). The algorithm is then applied to our problem and the three-loop correction is determined. 

\subsection{An exact recursion relation}
\la{sec:recursion}

Let us write (\ref{4.9}) as
\be
\la{6.3}
G(x) = \sum_{n=0}^{\infty}\frac{q^{n\Delta}}{(q;q)_{n}}\,Z_{n}(x)^{2}, \qquad
Z_{n}(x) \equiv \sum_{p=0}^{\infty}\frac{2(-1)^{p}}{x}\,q^{p+\binom{p}{2}}\,\frac{(q;q)_{n+p}}{(q;q)_{p}}\,(n+2p+1)\,I_{n+2p+1}(x).
\ee
Comparing $n=0$ of (\ref{6.3}) with (\ref{4.8}) gives 
$Z_{0}(x)=Z(x)$, \ie the partition function. Comparison with the first line of (\ref{3.11}) shows that
\be
\la{6.4}
Z_{n}(x) =(-1)^{n} (q;q)_{\infty}\int_{0}^{\pi}\frac{d\theta}{2\pi}\,e^{-x\cos\theta}\,|(e^{2i\theta};q)_{\infty}|^{2}\,H_{n}(\cos\theta|q),
\ee
where the sign factor, disappearing in (\ref{6.3}), is fixed by looking at the first non-zero contribution at small $x$.
The continuous $q$-Hermite polynomials (\ref{3.3}) satisfy the standard three-term recurrence
\be
\la{6.5}
H_{n+1}(\cos\mu|q) = 2\cos\mu\,H_{n}(\cos\mu|q) - (1-q^{n})\,H_{n-1}(\cos\mu|q), \qquad H_{-1}\equiv0,\ H_{0}\equiv1.
\ee
Since $\cos\theta\,e^{-x\cos\theta}=-\partial_{x}e^{-x\cos\theta}$, applying the transform (\ref{6.4}) to both sides of (\ref{6.5}) gives
\be
\la{6.6}
Z_{n+1}(x) = 2\,Z_{n}'(x) - (1-q^{n})\,Z_{n-1}(x), \qquad Z_{-1}(x)=0,\quad Z_{0}(x)=Z(x).
\ee
The construction is purely algebraic: each step requires only one differentiation and one linear combination. After expansion in small $x$, 
every $Z_{n}(x)$ produced by (\ref{6.6}) is a power series in $x$ with exact rational-in-$q$ coefficients. The substitution $q=e^{-\l}$ and 
the subsequent expansion in $\l$ 
should therefore be carried out only once, at the end of the calculation. This mirrors the direct approach of Section~\ref{sec:small-beta},
except that the recursion  (\ref{6.6}) replaces the considerably more expensive $q$-oscillator algebra.

\subsection{Three-loop result}

Using the Ansatz (\ref{6.1}) with $N_{1}=8$, $N_{2}=11$, and the degree of $\mc P_{p}$ equal to 5, we can compare the resulting Ansatz with the three-loop contribution expanded up to order $u^{180}$.
This greatly exceeds the number of coefficients required to fix the Ansatz and therefore provides a highly non-trivial validation of the procedure.
The explicit three-loop result reads
\ba
\la{6.7}
\mc A_{3}(u,0) &= -\frac{1}{1536 (1+u \tan u)^8}[
u^2 (-31+9 u^2)+u (55-146 u^2-39 u^4) \tan u \lp
+(-24+343 u^2-691 u^4+3 u^6) \tan ^2 u
+u (-131+1053 u^2-1281 u^4-63 u^6) \tan^3 u \lp
+(-24-486 u^2+2327 u^4-1503 u^6-36 u^8) \tan^4 u \lp
+u (-48-1364 u^2+3733 u^4-1290 u^6-12 u^8) \tan^5 u \lp
+(-24+120 u^2-2562 u^4+4152 u^6-668 u^8) \tan^6 u \lp
+u (-120+512 u^2-3125 u^4+3225 u^6-200 u^8) \tan^7 u \lp
-8 u^2 (30-91 u^2+310 u^4-221 u^6+3 u^8) \tan^8 u
+4 u^3 (-60+132 u^2-313 u^4+165 u^6) \tan^9 u \lp
+8 u^4 (-15+25 u^2-46 u^4+19 u^6) \tan^{10} u
+8 u^5 (-3+4 u^2-6 u^4+2 u^6) \tan^{11} u].
\ea
Its small $u$ expansion has initial terms
\ba
\mc A_{3}(u,0) &= -\frac{5 u^4}{144}+\frac{101 u^6}{240}-\frac{5639 
u^8}{2880}+\frac{2312143 u^{10}}{388800}-\frac{416424439 
u^{12}}{29937600}+\frac{6169176487 
u^{14}}{227026800}\lp
-\frac{5698759688849 u^{16}}{122594472000}+\frac{44881573998113 u^{18}}{625231807200}+\cdots,
\ea
reproducing minus the $O(\l^{3})$ terms in (\ref{4.3}).

To summarize the three closed forms obtained in this section and the previous one, 
Figure~\ref{fig:coefficients} shows $\mc A_{1,2,3}(u,0)$, rescaled by $\tan^{k}u$ to remain 
finite over the whole range, as functions of $u\in(0,\pi/2)$. Each curve vanishes at $u\to0$ 
(infinite temperature) and approaches a constant as $u\to\pi/2$ (low temperature) -- the 
coefficients $K_{1,2,3}$ already anticipated in (\ref{1.23})--(\ref{1.24}), and discussed in 
Section~\ref{sec:analysis} below.

\begin{figure}[htb]
\centering
\begin{subfigure}[t]{0.48\textwidth}
\centering
\includegraphics[width=\textwidth]{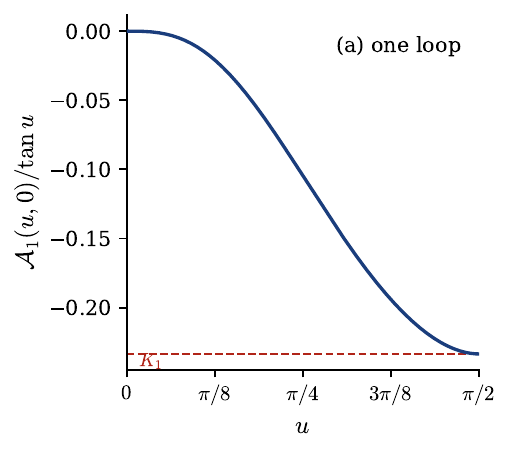}
\end{subfigure}
\hfill
\begin{subfigure}[t]{0.48\textwidth}
\centering
\includegraphics[width=\textwidth]{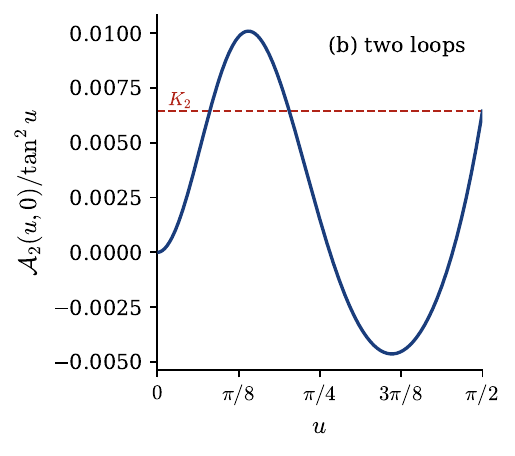}
\end{subfigure}

\vspace{0.5cm}

\begin{subfigure}[t]{0.48\textwidth}
\centering
\includegraphics[width=\textwidth]{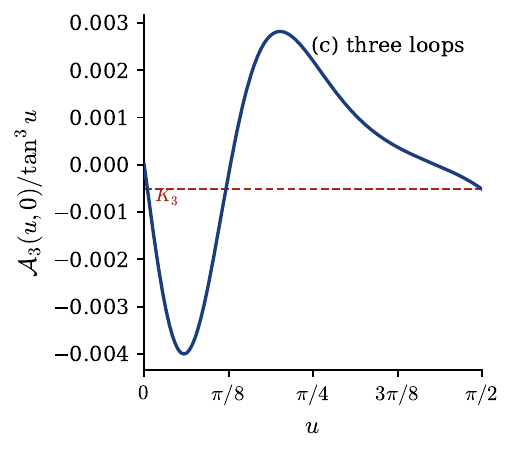}
\end{subfigure}
\hfill
\begin{subfigure}[t]{0.48\textwidth}
\phantom{\includegraphics[width=\textwidth]{A3_panel.pdf}}
\end{subfigure}
\caption{The rescaled loop coefficients $\mc A_{k}(u,0)/\tan^{k}u$, $k=1,2,3$, over the full 
temperature range $u\in(0,\pi/2)$. Each curve vanishes as $u\to0$ (infinite temperature) and 
approaches the constant $K_{k}$ (dashed line) as $u\to\pi/2$ (low temperature), consistently 
with (\ref{7.3}).}
\la{fig:coefficients}
\end{figure}

\section{Low-temperature regime and the Schwarzian limit}
\la{sec:analysis}

This section has two aims. The first is to derive the low-temperature expansion of $L_{0}$ 
from the exact expressions of the previous sections, and to show that it organizes into a 
double-scaling limit connecting DSSYK to JT gravity, with $\l\beta$ playing the role of the 
Schwarzian coupling. The second, in Section~\ref{sec:schwarzian-matching}, is to check the 
resulting one-loop coefficient against an independent computation performed entirely within 
the Schwarzian theory.

The results of the previous sections establish that
\be
\la{7.1}
L_{0} = -2\log\cos u-\mc A_{1}(u,0)\l-\mc A_{2}(u,0)\,\l^{2}-\mc A_{3}(u,0)\, \l^{3}+\cdots, 
\ee
where $\mc A_{1,2,3}(u,0)$ are given in (\ref{5.39}), (\ref{5.53}), and (\ref{6.7}), respectively, and $u$ is related to the inverse temperature $\beta$ by $\beta=\frac{4u}{\cos u}$
(at $J=1/2$). 
The low-temperature limit of $L_{0}$ corresponds to $u\to \pi/2$ and, more precisely:
\be
u = \frac{\pi}{2}\bigg(1-\frac{4}{\beta}+\frac{16}{\beta^{2}}-\frac{8(24+\pi^{2})}{4\beta^{3}}+\frac{128(6+\pi^{2})}{3\beta^{4}}-\frac{8(1920+800\pi^{2}+9\pi^{4})}{15\beta^{5}}+\cdots\bigg).
\ee
This gives
\ba
\la{7.3}
L_{0} &= 2\log\frac{\beta}{2\pi}+\frac{8}{\beta}-\frac{16}{\beta^{2}}+\cdots
+\bigg[\bigg(\frac{1}{16}-\frac{1}{4 \pi ^2}\bigg) \beta +\bigg(\frac{1}{4}-\frac{1}{\pi^2}\bigg)+\bigg(-\frac{3}{2}-\frac{\pi ^2}{8}\bigg)\frac{1}{\beta}+\cdots\bigg]\, \l \lp
+\bigg[\bigg(-\frac{1}{512}-\frac{1}{32 \pi ^4}+\frac{1}{48 \pi ^2}\bigg) \beta^2+\bigg(-\frac{1}{4 \pi ^4}+\frac{5}{48 \pi ^2}\bigg) \beta+\bigg(-\frac{3}{16}-\frac{1}{2 \pi ^4}+\frac{5}{12 \pi ^2}
+\frac{\pi^2}{128}\bigg)\beta^{0}+\cdots\bigg]\, \l^{2}\lp
+\bigg[\bigg(\frac{1}{6144}-\frac{1}{64 \pi ^6}+\frac{1}{192 \pi ^4}-\frac{1}{512 \pi^2}\bigg) \beta ^3+\bigg(\frac{1}{1024}-\frac{3}{16 \pi ^6}+\frac{3}{64 \pi^4}
-\frac{5}{384 \pi ^2}\bigg) \beta ^2\lp
+\bigg(\frac{23}{1536}-\frac{3}{4 \pi^6}
+\frac{7}{32 \pi ^4}-\frac{25}{384 \pi ^2}-\frac{\pi ^2}{1024}\bigg) 
\beta\bigg]\, \l^{3}+\cdots\ .
\ea
The leading large-$\beta$ contribution at each loop order $k$ is $(\l\beta)^{k}$: as already 
discussed in the Introduction, see (\ref{1.23}), this is the Schwarzian coupling of the 
triple-scaling limit connecting DSSYK to JT gravity \cite{Berkooz:2018jqr,Lin:2022rbf}, whose 
Schwarzian description was developed in \cite{Maldacena:2016upp,Jensen:2016pah,Engelsoy:2016xyb} 
(see \cite{Sarosi:2017ykf,Mertens:2022irh} for reviews). The dictionary can be made explicit.
The low temperature saddle approaches the 
edge of the spectrum, $\theta_{*}=\pi/2-u\to0$, where the appropriate scaling variable is 
$\theta=\l\kappa$ with $\kappa$ fixed as $\l\to0$ \cite{Berkooz:2018jqr,Lin:2022rbf}. 
The energy 
(\ref{3.1}) measured from the edge is $E(\theta)-E(0)=-J\l\kappa^{2}+O(\l^{2})$, so that the 
Boltzmann weight becomes $e^{-J\l\beta\kappa^{2}}=e^{-\mk b\kappa^{2}}$, with $\mk b$ given by 
\be
\la{7.4}
\mk b \equiv \frac{\beta_{\rm Sch}}{2C} = J\,\l\beta = \frac{\l\beta}{2},
\ee
where $C$ is the Schwarzian coupling.
The spectral 
measure  in (\ref{3.5}) reduces to the Schwarzian density of states 
\cite{Maldacena:2016upp,Mertens:2017mtv},
\be
\la{7.5}
(q;q)_{\infty}\,|(e^{2i\theta};q)_{\infty}|^{2}\Big|_{\theta=\l\kappa} \ \propto\ 
\kappa\,\sinh(2\pi\kappa)\,\big[1+O(\l)\big].
\ee
Finally, the classical saddle of Section~\ref{sec:saddle}, \cf \cite{Goel:2023svz,Okuyama:2023bch}, 
maps onto the Schwarzian one: from $\beta=4u/\cos u$ one has $\theta_{*}\to2\pi/\beta$ as 
$u\to\pi/2$, and thus
\be
\la{7.6}
\kappa_{*} = \frac{\theta_{*}}{\l} = \frac{2\pi}{\l\beta} = \frac{\pi}{\mk b},
\ee
which is of course the saddle point of $\kappa\sinh(2\pi\kappa)\,e^{-\mk b\kappa^{2}}$. 

This provides a check of (\ref{7.1}) at the classical level, and clarifies the origin of the 
additive constant. Indeed, since $-2\log\cos u=-2\log\sin\theta_{*}\to-2\log\theta_{*}$, we have
\be
\la{7.7}
L_{0}\big|_{\rm class} = -2\log\kappa_{*}+2\log\frac{1}{\l},
\ee
while the Schwarzian evaluation of the two-sided geodesic length gives $-2\log\kappa_{*}$ up to 
an additive constant. The two agree, and the leftover $2\log(1/\l)$ is not a discrepancy: it is 
the divergent renormalization that must be subtracted to pass from the chord number to the JT 
geodesic length \cite{Lin:2022rbf,Blommaert:2024whf}, and that appears in the sine-dilaton 
dictionary as the regularization length $\ell_{*}(\theta)$ \cite{Aguilar-Gutierrez:2025pqp}. 
It is scheme dependent on the gravity side and unambiguous in DSSYK, which is the point 
anticipated in the Introduction.

In terms of $\mk b$, the low temperature expansion reads 
$L_{0}=2\log\frac{1}{\l}+2\log\frac{\mk b}{2\pi}+\sum_{k\ge1}\wt K_{k}\,\mk b^{k}+\cdots$. 
Reading off the coefficient of the leading power of $\beta$ in each bracket of (\ref{7.3}) above 
and rescaling by $2^{k}$ gives $\wt K_{k}$, \ie
\bea
\la{7.8}
\wt K_{1} &= \frac{1}{8}-\frac{1}{2\pi^{2}}, \qquad
\wt K_{2} = -\frac{1}{128}+\frac{1}{12\pi^{2}}-\frac{1}{8\pi^{4}}, \\
\wt K_{3} &= \frac{1}{768}-\frac{1}{64\pi^{2}}+\frac{1}{24\pi^{4}}-\frac{1}{8\pi^{6}}.
\eea
The same coupling $\mk b$ organizes the low temperature expansion of the DSSYK partition 
function \cite{Beccaria:2026ndg}. A structural feature of (\ref{7.8}) is that each 
$\wt K_{k}$ is a polynomial of degree exactly $k$ in $1/\pi^{2}$ -- a fact we have verified 
directly at $k=1,2,3$ from the closed forms (\ref{5.39}), (\ref{5.53}), (\ref{6.7}).

\subsection{Matching the Schwarzian at one-loop}
\la{sec:schwarzian-matching}

The coefficients $\wt K_{k}$ 
in (\ref{7.8}) are the loop corrections to the two-sided geodesic length in JT 
gravity at the symmetric point, obtainable from the exact Schwarzian bilocal correlator 
\cite{Mertens:2017mtv,Kitaev:2018wpr}, reviewed in \cite{Mertens:2022irh}, by setting 
$\tau_{1}=0,\tau_{2}=\beta_{\rm Sch}/2$ and taking $-\partial_{\Delta}$ at $\Delta=0$, as in 
(\ref{1.6}).\footnote{The gravitational machinery required for such a computation is available 
at one loop in the exact sine-dilaton dual: the one-loop correction to the boundary-to-boundary 
propagator of the non-minimally coupled matter field dual to the DSSYK matter correlator was 
computed in \cite{Bossi:2024ffa} and shown to reproduce the corresponding quantum correction on 
the DSSYK side.}

Writing $\Gamma(\Delta\pm ix)\equiv\Gamma(\Delta+ix)\Gamma(\Delta-ix)$, one has, 
up to an overall $\Delta$-independent normalization, 
\ba
\la{7.9}
\langle & \mc O_{\Delta}(0)\mc O_{\Delta}(\beta_{\rm Sch}/2)\rangle \ \propto\lp
 \int_{0}^{\infty}\!dk_{1}^{2}\,dk_{2}^{2}\ 
\sinh(2\pi k_{1})\sinh(2\pi k_{2})\, e^{-\frac{\beta_{\rm Sch}}{4C}(k_{1}^{2}+k_{2}^{2})}\,
\frac{\Gamma(\Delta\pm i(k_{1}+k_{2}))\,\Gamma(\Delta\pm i(k_{1}-k_{2}))}{(2C)^{2\Delta}\,\Gamma(2\Delta)}.
\ea
Our goal is now to expand this quantity in large $C$ and we again use saddle point methods. \footnote{
A naive expansion of the Gaussian weight in powers of $1/C$ fails already at 
leading order: the density of states $\sinh(2\pi k_{1})\sinh(2\pi k_{2})$ grows exponentially in 
$k_{1},k_{2}$, and the Gaussian factors are needed for convergence.}
In the variables $\sigma=k_{1}+k_{2}$, $\delta=k_{1}-k_{2}$ the matrix element factorizes. Two 
further simplifications follow from the fact, established below by (\ref{7.12}), that the 
integral is dominated by $\sigma$ near its saddle $\sigma_{*}\sim C$ while $\delta=O(1)$. First, 
the density of states, exactly $\sinh(2\pi k_{1})\sinh(2\pi k_{2})=\tfrac12[\cosh(2\pi\sigma)-
\cosh(2\pi\delta)]$, is well approximated by treating each $k_{i}=(\sigma\pm\delta)/2$ as 
individually large, giving $\sinh(2\pi k_{1})\sinh(2\pi k_{2})\approx\tfrac14e^{2\pi\sigma}$ with 
no surviving $\delta$-dependent piece. Second, the exact measure $dk_{1}^{2}dk_{2}^{2}=
\tfrac12(\sigma^{2}-\delta^{2})\,d\sigma\,d\delta$ reduces to $\tfrac12\sigma^{2}\,d\sigma\,d\delta$, 
the $\delta^{2}$ term being a relative $O(\delta^{2}/\sigma^{2})\sim O(1/C^{2})$ correction, 
i.e.\ two-loop. One finds
\be
\la{7.10}
\langle\mc O_{\Delta}(0)\mc O_{\Delta}(\beta_{\rm Sch}/2)\rangle \ \propto\ \int_{0}^{\infty}\! d\sigma\int_{-\infty}^{\infty}\!d\delta\ 
\sigma^{2}\,e^{2\pi\sigma-\frac{\beta_{\rm Sch}}{8C}\sigma^{2}}\,e^{-\frac{\beta_{\rm Sch}}{8C}\delta^{2}}\,
\frac{\Gamma(\Delta\pm i\sigma)\,\Gamma(\Delta\pm i\delta)}{(2C)^{2\Delta}\Gamma(2\Delta)}.
\ee
To compute the $\delta$ integral, we expand the Gaussian weight in (\ref{7.10}) for large $C$,
\be
e^{-\frac{\beta_{\rm Sch}}{8C}\delta^{2}} = 1 - \frac{\beta_{\rm Sch}}{8C}\delta^{2} + O(1/C^{2}),
\ee
and use the Fourier transform \footnote{This follows from Fourier inversion of the transform $\int_{-\infty}^{\infty}dx\ e^{i\omega x}\sech^{2a}x = 2^{2a-1}\Gamma(a+i\omega/2)\Gamma(a-i\omega/2)/\Gamma(2a)$.}
\be
\la{7.12}
\frac{1}{\Gamma(2\Delta)}\int_{-\infty}^{\infty} \Gamma(\Delta\pm i\delta)\,e^{i\alpha\delta}\,d\delta \ =\ 
\frac{2\pi}{\big(2\cosh(\alpha/2)\big)^{2\Delta}} = \pi\, 2^{1-2\Delta}\, \bigg(1-\frac{\Delta}{4}\alpha^{2}+O(\alpha^{4})\bigg).
\ee
This gives
\be
\la{7.13}
\frac{1}{\Gamma(2\Delta)}\int_{-\infty}^{\infty} d\delta\,e^{-\frac{\beta_{\rm Sch}}{8C}\delta^{2}}\,\Gamma(\Delta\pm i\delta) 
\ =\ \pi\, 2^{1-2\Delta}\, \bigg(1-\frac{\beta_{\rm Sch}\Delta}{16\,C}+\cdots\bigg).
\ee
The next piece is the $\sigma$ integral, \ie
\be
\la{7.14}
\int_{0}^{\infty}d\sigma\, \sigma^{2}\, e^{2\pi \sigma-\frac{\beta_{\rm Sch}}{8C}\sigma^{2}}\Gamma(\Delta\pm i\sigma).
\ee
This cannot be treated the same way: with the density of states approximated 
as $e^{2\pi\sigma}$ and $\Gamma(\Delta\pm i\sigma)\sim e^{-\pi\sigma}$ at large $\sigma$, the net 
growth $e^{\pi\sigma}$ means the integral is finite only because of the Gaussian weight, not in 
spite of it -- the opposite of the $\delta$-direction, where $\Gamma(\Delta\pm i\delta)$ alone 
already guarantees convergence. Expanding the Gaussian is therefore not an option here; the 
Gaussian must instead be kept exactly and the integral evaluated at its saddle point.

We use 
$\Gamma(\Delta\pm i\sigma)\sim2\pi\sigma^{2\Delta-1}e^{-\pi\sigma}$ for $\sigma\to\infty$, turning 
(\ref{7.14}) into $2\pi\int_{0}^{\infty}d\sigma\,\sigma^{2\Delta+1}e^{\pi\sigma-\frac{\beta_{\rm Sch}}{8C}\sigma^{2}}$. 
This Laplace integral is dominated by its saddle point, at
\be
\sigma_{*} = \frac{4\pi C}{\beta_{\rm Sch}}, \qquad \text{consistently with } \kappa_{*}=\frac\pi{\mk b} 
\text{ in (\ref{7.6})}.
\ee
Expanding the power-law prefactor $\sigma^{2\Delta+1}$ to quadratic order in the Gaussian 
fluctuation around $\sigma_{*}$ gives
\be
\la{7.16}
\int_{0}^{\infty}d\sigma\,\sigma^{2}\,e^{2\pi\sigma-\frac{\beta_{\rm Sch}}{8C}\sigma^{2}}\Gamma(\Delta\pm i\sigma) 
\ =\ 2\pi\,\sigma_{*}^{2\Delta+1}\,e^{\frac{2\pi^{2}C}{\beta_{\rm Sch}}}\,\sqrt{\frac{8\pi C}{\beta_{\rm Sch}}}\,
\bigg(1+\frac{\beta_{\rm Sch}\,\Delta(2\Delta+1)}{4\pi^{2}C}+\cdots\bigg).
\ee
Assembling (\ref{7.16}) with the $\delta$-integral (\ref{7.13}) and the overall prefactor 
$(2C)^{-2\Delta}$ of (\ref{7.10}), and taking $-\partial_{\Delta}$ at $\Delta=0$: the 
$\Delta$-independent and $\Delta$-linear pieces of the three power-law factors 
$(2C)^{-2\Delta}$, $\sigma_{*}^{2\Delta}$, $2^{-2\Delta}$ combine into the classical term 
$2\log(\beta_{\rm Sch}/\pi)$ of (\ref{7.7}), while the two brackets contribute 
$-\frac{\beta_{\rm Sch}}{4\pi^{2}C}$ (from the $\sigma$ saddle) and $+\frac{\beta_{\rm Sch}}{16C}$ 
(from the $\delta$ integral). This gives the length at one loop,
\be
\la{7.17}
\langle L\rangle_{\rm Sch} \ =\ 2\log\frac{\beta_{\rm Sch}}{\pi} + \frac{\beta_{\rm Sch}}{C}
\bigg(\frac{1}{16}-\frac{1}{4\pi^{2}}\bigg) + O(1/C^{2}),
\ee
where the rational $1/16$ 
part of the one-loop correction comes from the $\delta$ integral (\ref{7.13}), and the $-1/(4\pi^{2})$ term from fluctuations around the $\sigma$ saddle
(\ref{7.16}).
 
With the dictionary $C=1/\l$ fixed in (\ref{7.4}) and $\beta_{\rm Sch}=\beta$, 
(\ref{7.17}) reproduces $\wh K_{1}\,\l\beta$ of (\ref{1.24}) exactly -- both the rational and the 
$\pi$-dependent pieces -- providing an independent confirmation, from the Schwarzian description, of 
$\mc A_{1}(u,0)$ in the strict $u\to\pi/2$ limit. 
\footnote{A further, 
independent check comes from a genuinely different method: \cite{Okuyama:2023bch} (Section 
6.1) compute the one-loop correction to the two-point function at arbitrary operator 
separation directly within Schwarzian quantum mechanics, via reparametrization-mode 
perturbation theory with the propagator of \cite{Maldacena:2016upp}, rather than through the 
exact matrix element (\ref{7.9}) used here. 
Identifying their mode coupling with $\kappa_{*}$ of (\ref{7.6}), their result agrees with 
$\l\mc A_{1}(u,\phi)$ identically, not only at $\phi=0$, and in particular reproduces $\wh K_{1}$.} 
The residual mismatch in the additive constant, 
$2\log(\beta/\pi)$ here against $2\log(\beta/2\pi)$ in (\ref{7.7}), is exactly the scheme-dependent 
renormalization already discussed in the Introduction. 

The subleading-in-$1/\beta$ terms at each fixed loop order $k$ in (\ref{7.3}) are a separate 
matter: they have no counterpart in this calculation, which by construction produces only the 
leading $\beta^{k}$ piece of each bracket, cf.\ (\ref{7.17}). 
They encode genuine sine-dilaton physics invisible to pure JT gravity: since sine-dilaton 
gravity is the exact dual of DSSYK at any $\l$, and the strict Schwarzian limit computed above 
reproduces only the leading $\beta^{k}$ piece, the remaining terms must originate in the 
finite-$\l$ completion of the dictionary, of the kind matched directly against DSSYK at one 
loop in \cite{Bossi:2024ffa}.

\section{Fluctuations of the wormhole length at $t=0$}
\la{sec:fluctuations}

\subsection{Variance}

A natural companion to $L_{0}$ is the variance of the length operator at the same symmetric 
point,
\be
\la{8.1}
L_{0}^{(2)}(\beta) \equiv \langle\hat\ell^{2}\rangle_{c}\big|_{t=0} = \langle\hat\ell^{2}\rangle-\langle\hat\ell\rangle^{2},
\ee
quantifying the size of quantum fluctuations of the wormhole length around its mean value $L_{0}$. 
Such fluctuations, and higher cumulants more generally, have been studied in the holographic 
context in \cite{Iliesiu:2021ari,Fu:2025kkh}. At infinite temperature, the semiclassical expansion of 
$L_{0}^{(2)}$ was obtained to four-loop order in \cite{Alfinito:2026cky}; 
here we extend the computation to arbitrary temperature, still at $t=0$, using the same 
methods developed in the previous sections.

Since $\hat\ell=\l\wh n$, one has directly $L_{0}^{(2)}=\l^{2}(\langle\wh n^{2}\rangle-
\langle\wh n\rangle^{2})$ and  
the cumulant expansion of (\ref{1.9}) gives 
\be
\la{8.2}
L_{0}^{(2)} = 2\mc B-\mc A^{2},
\ee
where $\mc A,\mc B$ are the coefficients of $\Delta,\Delta^{2}$ in (\ref{1.9}). 
The loop corrections are
\be
\la{8.3}
L_{0}^{(2)} = \l\,\mc V_{1}+\l^{2}\,\mc V_{2}+\l^{3}\,\mc V_{3}+\cdots.
\ee
At one loop, using 
$\mc B_{1}(u,0)$ from (\ref{1.11}) at $\zeta=0$,
\be
\la{8.4}
\mc B_{1}(u,0) = \frac{\tan^{2}u}{2(1+u\tan u)}, \qquad\text{so that}\qquad
\mc V_{1} = \frac{\tan^{2}u}{1+u\tan u}.
\ee
This one-loop coefficient is manifestly positive, as it should,  throughout the physical range $u\in(0,\pi/2)$.

As for $L_{0}$ itself, the $q$-oscillator algebra of Section~\ref{sec:small-beta} gives access to 
$L_{0}^{(2)}$ order by order in $u$, exactly in $\l$. Writing $\langle\wh n^{2}\rangle=\sum_{j}h_{j}(q)(J\beta/\sqrt\l)^{2j}$,
paralleling (\ref{4.2}), the first three coefficients are
\be
\la{8.5}
h_{1}=\frac14, \qquad h_{2}=\frac{4q-1}{48}, \qquad h_{3}=\frac{39q^{3}+97q^{2}-161q+25}{5760},
\ee
from which (\ref{8.1}) follows, exact in $\l$, to the same order:
\be
\la{8.6}
L_{0}^{(2)} = \l\,u^{2} + \l\bigg[1+\frac{4(q-1)}{3\l}\bigg]u^{4} 
+ \l\bigg[\frac{2}{3}+\frac{8(q-1)}{3\l}+\frac{107(q-1)^{2}}{45\l^{2}}+\frac{13(q-1)^{3}}{30\l^{2}}\bigg]\,u^{6}+\cdots.
\ee
Expanding in $\l$ reproduces (\ref{8.4}) at one loop, and gives the corresponding two-loop 
coefficients for free. Notably, the $u^{2}$ term receives no correction beyond one loop.

Longer expansions in $u$, needed to fix an Ansatz analogous to (\ref{1.19}), are obtained exactly 
as in Section~\ref{sec:recursion}: the same exact recursion relation for $Z_{n}(x)$ applies 
unchanged, since it follows from the $q$-Hermite recurrence alone and does not depend on which 
moment of $n$ is being computed; only the final weighted sum over $n$ differs by the substitution $n\to n^{2}$.

Using this strategy, the two-loop coefficient of $L_{0}^{(2)}$ is found to be
\ba
\la{8.7}
\mc V_{2} &= \frac{1}{8(1+u\tan u)^{4}}\bigg[2u\tan u+(-2+14u^{2})\tan^{2}u+u(-7+20u^{2})\tan^{3}u\lp
+(-1-9u^{2}+15u^{4})\tan^{4}u
+u(-2-6u^{2}+6u^{4})\tan^{5}u+(-u^{2}-2u^{4}+u^{6})\tan^{6}u\bigg],
\ea
obtained from a structural Ansatz of the same kind as (\ref{1.19}). 

The small $u$ expansions of $\mc V_{1}$ and $\mc V_{2}$ read
\be
\mc V_{1} = u^2 - \frac{u^4}{3} + \frac{17 u^6}{45} - \frac{64 u^8}{315} +\frac{2342 u^{10}}{14175}+\cdots, \qquad
\mc V_{2} = \frac{2u^{4}}{3}-\frac{133u^{6}}{90}+\frac{683u^{8}}{280}-\frac{185077u^{10}}{56700}+\cdots.
\ee
The low-temperature expansion, similar to (\ref{7.3}), reads
\ba
\la{8.9}
L_{0}^{(2)} &=
\bigg(\frac{\beta}{\pi^{2}}+\frac{4}{\pi^{2}}-\frac{2}{\beta}+\cdots\bigg)\, \l \lp
+\bigg[\bigg(\frac{1}{128}-\frac{1}{8\pi^{4}}-\frac{1}{16\pi^{2}}\bigg) \beta^2+\bigg(\frac{1}{16}-\frac{1}{\pi^{4}}-\frac{1}{4\pi^{2}}\bigg) \beta+
\bigg(\frac{3}{8}-\frac{2}{\pi^{4}}-\frac{1}{\pi^{2}}-\frac{\pi^{2}}{32}\bigg)\beta^{0}+\cdots\bigg]\, \l^{2}+\cdots\ .
\ea
The one-loop term $\frac{\beta\l}{\pi^{2}}$ is easily checked by our previous computation in the Schwarzian theory, the relevant term being the last factor in  (\ref{7.16})
\be
\la{8.10}
\partial_{\Delta}^{2}\log\bigg(1+\frac{\beta_{\rm Sch}}{4\pi^{2}C}\Delta(2\Delta+1)+\cdots\bigg)\bigg|_{\Delta=0} = \frac{1}{\pi^{2}}\frac{\beta_{\rm Sch}}{C}+\cdots,
\ee
where differentiation of the logarithm of the 2-point function gives the cumulant and we dropped terms whose logarithm is only linear in $\Delta$.

\subsection{Third-order cumulant}

The same kind of analysis can be repeated for the third-order cumulant of the length operator at the symmetric 
point,
\be
\la{8.11}
L_{0}^{(3)}(\beta) \equiv \langle\hat\ell^{3}\rangle_{c}\big|_{t=0} = \langle\hat\ell^{3}\rangle-3\langle\hat\ell^{2}\rangle\,\langle\hat\ell\rangle+2\langle\hat\ell\rangle^{3}.
\ee
This quantity requires the third logarithmic derivative of the two-point function with respect to $\Delta$ 
and cannot be obtained from (\ref{1.9}), not even at leading order, because the $O(\Delta^{3})$ is not available. Here, we just look at the leading 
two-loop correction $\mc C_{2}$ in 
\be
\la{8.12}
L_{0}^{(3)} = \l^{2}\,\mc C_{2}+\l^{3}\,\mc C_{3}+\cdots.
\ee
Again, the $q$-oscillator algebra of Section~\ref{sec:small-beta} gives access to 
$L_{0}^{(3)}$ order by order in $u$, exactly in $\l$. We find \footnote{
The results above, and their generalization to arbitrary order, follow most directly by 
working with the generating function $\langle e^{zn}\rangle$ rather than with individual 
moments: this is accessible in closed form by the same $q$-oscillator algebra, and the 
$p$-th cumulant follows from $\partial_{z}^{p}\log\langle e^{zn}\rangle|_{z=0}$, valid for any $p$. 
Proceeding this way, one finds, for a generic $p$-th order cumulant, the formula 
\ba
\la{8.13}
L_{0}^{(p)} &= \l^{p-1}u^{2}+\l^{p-1}\bigg[1+\frac{(4+3\cdot 2^{p})(q-1)}{12\l}\bigg]\,u^{4}+\l^{p-1}\bigg[
\frac{2}{3}
+\frac{(4+3\cdot 2^{p})(q-1)}{6\l} \lp
+\frac{(23+45\cdot 2^{p}+25\cdot 3^{p})(q-1)^{2}}{180\l^{2}}
+\frac{(6+15\cdot 2^{p}+10\cdot 3^{p})(q-1)^{3}}{360\l^{2}}\bigg]\,u^{6}+\cdots , 
\ea
which reproduces the $p=1, 2, 3$ cases in (\ref{4.4}, \ref{8.6}, \ref{8.14}). }
\be
\la{8.14}
L_{0}^{(3)} = \l^{2}\,u^{2} + \l^{2}\bigg[1+\frac{7(q-1)}{3\l}\bigg]\,u^{4} 
+ \l^{2}\bigg[\frac{2}{3}+\frac{14(q-1)}{3\l}+\frac{529(q-1)^{2}}{90\l^{2}}+\frac{11(q-1)^{3}}{10\l^{2}}\bigg]\,u^{6}+\cdots.
\ee
Using the same long high-temperature expansions, we obtain the exact expression
\ba
\la{8.15}
\mc C_{2} &= \frac{8\tan^{2}u-4u\tan^{3}u+12(1-u^{2})\tan^{4}u+4u(3-u^{2})\tan^{5}u}{8(1+u\tan u)^{3}}.
\ea
The small $u$ expansion of $\mc C_{2}$ reads
\be
\la{8.16}
\mc C_{2} = u^2 - \frac{4 u^4}{3} + \frac{169 u^6}{90} - \frac{652 u^8}{315} +
\frac{61819 u^{10}}{28350} -\frac{197017 u^{12}}{93555}+\cdots.
\ee
The low-temperature expansion, \cf  (\ref{7.3}), following from the expansion of (\ref{8.15}) around $u=\pi/2$ is 
\ba
\la{8.17}
L_{0}^{(3)} &=\bigg[\bigg(\frac{3}{2\pi^{4}}-\frac{1}{8\pi^{2}}\bigg)\, \beta^{2}
+\bigg(\frac{12}{\pi^{4}}-\frac{1}{\pi^{2}}\bigg)\, \beta+\bigg(\frac{1}{2}+\frac{24}{\pi^{4}}-\frac{4}{\pi^{2}}\bigg)\, \beta^{0}+\cdots\bigg]\, \l^{2}+\cdots.
\ea

\section{Summary and comments}
\la{sec:summary}

In this paper, we computed the onset of the wormhole length, $L_{0}$, in the thermofield double state of 
the DSSYK model, \ie the Krylov spread complexity of the zero-chord state after purely Euclidean 
evolution, in the semiclassical limit and at arbitrary temperature, through three loops.
The classical  and  one-loop terms were already known, whereas the two- and 
three-loop coefficients in (\ref{5.53}) and (\ref{6.7}) are new.

The calculation has been organized along three independent routes. The first is a direct 
evaluation of the defining matrix elements by means of the $q$-oscillator algebra, which is 
exact in the semiclassical coupling $\l$ but limited to relatively low order in the high temperature expansion. 
The second is a saddle point analysis of the DSSYK two-point function at coincident insertion 
points, which is exact in the temperature but becomes increasingly involved at higher loop order; we 
have pushed it to two loops. The third approach combines the recursion relation (\ref{6.6}) with the 
structural Ansatz (\ref{6.1}), trading the saddle point expansion for a long but purely 
algebraic high temperature expansion, and this is what gives access to three loops. The three 
methods agree wherever they overlap, and the last of them is validated on a number of data 
points far in excess of what is needed to fix the Ansatz.
In principle, given a sufficiently 
long high-temperature expansion, the same strategy extends straightforwardly to four and 
higher loop order.

Two features of the result deserve to be emphasized. The first is structural: all the 
coefficients $\mc A_{k}(u,0)$ are rational in $u$ and $\tan u$, with the degrees 
(\ref{6.2}) growing linearly with the loop order. This is what makes the Ansatz strategy 
viable, and it also predicts the shape of the four-loop coefficient. The second concerns the 
role of the singularities at coincident points. At each loop order the separate ingredients of 
the fluctuation determinant are singular in the regulator $\eps$, with the order of the poles 
increasing with the loop order, and the finiteness of $L_{0}$ results from a cancellation among 
them that we have exhibited explicitly at one and two loops. This is a feature specific to the symmetric 
point $t=0$, and it may be one reason why this value of the complexity is usually 
subtracted rather than computed.

The same strategy extends beyond $L_{0}$ itself: in Section~\ref{sec:fluctuations} we 
compute the variance and third-order cumulant of the length at the same symmetric point, through two loops, showing that 
the methods developed in this paper are not specific to the mean length but apply equally to its 
fluctuations.

We conclude by addressing the question of whether it is worthwhile to compute a quantity whose bulk dual is conventionally subtracted.
We believe this is useful for at least two reasons. First, the divergence removed on 
the gravity side is $\beta$-independent: by (\ref{7.7}), only the additive constant 
$2\log(1/\l)$ is scheme dependent, while the entire functional dependence on temperature -- the 
classical piece $-2\log\kappa_{*}(\beta)$ and every loop correction $\mc A_{k}(u,0)$ computed in 
this paper -- is untouched by whatever regularization is chosen on the gravity side. 
This is not merely a matter of convention. A consistent holographic renormalization fixes the boundary cutoff once and for all, 
independently of the state under consideration. Consequently, rescaling the cutoff can only shift the answer by a temperature-independent constant.
Subtracting the full function $L_{0}(\beta)$ would instead require a different 
prescription at every temperature, which is not holographic renormalization and 
would conflict with treating the bulk-to-boundary map as a single 
dictionary valid for every state.

Second, and more fundamentally, DSSYK has no such ambiguity to begin with: $\hat\ell=\l\hat n$ 
is a self-adjoint operator with a discrete, bounded-below spectrum, and 
$L_{0}(\beta,\l)=\langle\hat\ell\rangle_{\beta}$ is an unambiguous finite number at every 
temperature and every loop order. 
Quantities defined as pure differences, such as 
$\langle\ell(t)\rangle-\langle\ell(0)\rangle$ in \cite{Iliesiu:2021ari}, are of course 
insensitive to $L_{0}$ by construction, and this is a different, narrower question than the one 
addressed here; but the length of the Einstein-Rosen bridge at rest, as a function of 
temperature, is a genuine observable, and the present results fix it unambiguously, order by 
order in $\l$.

This last point can be sharpened into a concrete benchmark for future work. Any computation of 
the finite-temperature complexity $C_{\beta}(t)$ that extends the infinite-temperature, all-$t$ 
results of \cite{Alfinito:2026cky} to generic $\beta$ must, at $t=0$, reduce order by order in 
$\l$ to the coefficients $\mc A_{1,2,3}(u,0)$ obtained here, which is a nontrivial requirement.

\section*{Acknowledgements}

We thank S. E. Aguilar-Gutierrez for useful comments.
MB is supported by the INFN grant GAST. EA is supported by the MUR project GINEVRA, prot.
2022BZYBWM.

\bibliography{Krylov-Biblio}
\bibliographystyle{JHEP-v2.9}
\end{document}